\documentclass[twocolumn]{article}
\usepackage{amsmath,amssymb,amsfonts}
\usepackage{cuted}  % Add the cuted package
\usepackage{caption}
\usepackage{algorithmic}
\usepackage{booktabs}
\usepackage{graphicx}
\usepackage{textcomp}
\usepackage{xcolor}
\usepackage{tabularx, multirow}
\usepackage{hyperref} % Add this line to enable hyperlinks
\usepackage{fancyhdr,lipsum}
\usepackage{subcaption}
\usepackage[shortcuts]{extdash}

\usepackage[%
backend=biber,bibencoding=utf8, %instead of bibtex
language=auto,
style=ieee,
sorting=none, % nyt for name, year, title
maxbibnames=10, % default: 3, et al.
natbib=true % natbib compatibility mode (\citep and~\citet still work)
]{biblatex}
\bibliography{report.bib}

\hypersetup{
colorlinks=true,
    linkcolor=black, % Change the link color to blue
    urlcolor=blue, % Change the URL color to blue
    citecolor=black, % Change the citation color to blue
    filecolor=black % Change the file color to blue
}

\def\app#1#2{%
  \mathrel{%
    \setbox0=\hbox{$#1\sim$}%
    \setbox2=\hbox{%
      \rlap{\hbox{$#1\propto$}}%
      \lower1.1\ht0\box0%
    }%
    \raise0.25\ht2\box2%
  }%
}

\newcommand{\carbon}{CO$_2$}
\newcommand{\hydrogen}{H$_2$}
\newcommand{\carbongrid}{CO$_2$\=/Grid}
\newcommand{\hydrogengrid}{H$_2$\=/Grid}
\newcommand{\baselinescenario}{Baseline scenario}
\newcommand{\carbonscenario}{CO$_2$\=/Grid scenario}
\newcommand{\hydrogenscenario}{H$_2$\=/Grid scenario}
\newcommand{\hybridscenario}{Hybrid scenario}

\graphicspath{
    {results/}
}

\begin{document}

% FORMATTING:
% - https://www.nature.com/nenergy/submission-guidelines/aip-and-formatting
% - https://www.nature.com/nenergy/content abstract: 150 words, main text: 3000 words, figures: 8

\title{\hydrogen{} and \carbon{} Network Strategies for the European Energy System}
% \title{Competition and Synergies of \hydrogen{} and \carbon{} Networks in Europe}

% \title{Synthetic fuels in Europe: Transport Hydrogen to Carbon, or Carbon to Hydrogen?}

\author{
    Fabian Hofmann, Christoph Tries, Fabian Neumann, Elisabeth Zeyen, Tom Brown \\
    \textit{Institute of Energy Technology} \\
    \textit{Technical University of Berlin}\\
    Berlin, Germany \\
    m.hofmann@tu-berlin.de
}

\maketitle

\begin{abstract}
    Hydrogen and carbon dioxide transport can both play an essential role in climate-neutral energy systems. Hydrogen networks help serve regions with high energy demand, while excess emissions are transported away in carbon dioxide networks. For the synthesis of carbonaceous fuels, it is less clear which input should be transported: hydrogen to carbon point sources or carbon to low-cost hydrogen. We explore both networks' potential synergies and competition in a cost-optimal carbon-neutral European energy system. In a direct comparison, a hydrogen network is more cost-effective than a carbon network, as it serves to transport hydrogen to demand and to point source of carbon for utilization. However, in a hybrid scenario where both networks are present, the carbon network effectively complements the hydrogen network, promoting carbon capture from distributed biomass and reducing reliance on direct air capture. The layouts of the hydrogen and carbon dioxide networks are robust if the climate target is tightened to be net-negative.
\end{abstract}

\section{Introduction}

The transition to a climate-neutral European economy is a pressing challenge that demands coordinated action across various energy sectors. While management of both carbon dioxide (\carbon{}) and hydrogen (\hydrogen{}) is considered a critical component of this transition, a gap exists in understanding how new hydrogen infrastructure effectively interacts with carbon management technologies, including carbon capture, transport, storage, utilization, and sequestration. Hydrogen is being considered in several industries that are not eligible for electrification. It offers an efficient way to transport and store energy over long distances and periods and can be produced from electrolysis or steam methane reforming (SMR), for example. Captured carbon can be stored in geological formations permanently, a process known as carbon sequestration (CS). Additionally, carbon can be combined with hydrogen to produce carbonaceous fuels, such as synthetic kerosene for aviation, synthetic methanol for shipping or synthetic methane for industrial feedstock or synthetic naphtha as a feedstock for high value chemicals. The production of these carbonaceous fuels is jointly referred to as Carbon Utilization (CU). Carbon can be effectively captured from industrial processes and the combustion of biomass, fossil fuels, or synthetic carbonaceous fuels through point-source carbon capture (CC) techniques, or harvested from the atmosphere using direct air capture (DAC). Both \carbon{} and \hydrogen{} networks are likely to play a crucial role in the cost-effective integration of these carbon management technologies to enable net-zero economies.

% Recently, policymakers in Europe have started developing hydrogen and carbon management strategies,
% ~\cite{GermanyDevelopingStrategy2023,CarbonManagementStrategie}, planning infrastructure components~\cite{CONetz}, and committing to the first carbon utilization projects~\cite{EFuelsPilotPlant2022,OrstedAssumesFull,GROUNDBREAKINGEFUELPRODUCTION,DLREfuelsDLR}.
In line with the climate neutrality target by 2050 of the European Union (EU) under the European Green Deal~\cite{europeangreendeal}, numerous programs, financial models, and initiatives have been launched to support the ramp-up of the hydrogen and carbon economy. The Renewable Energy Directive III~\cite{DirectiveEU20232023} aims to produce 10~Mt hydrogen from renewable energy sources domestically in the EU. The Net-Zero Industry Act~\cite{eu2023netzero} proposes a \carbon{} injection target of 50~Mt/a by 2030, which will be sequestered within the European Union. The European Commission's 2024 industrial carbon management strategy~\cite{comcms} plans for capture of 450~Mt/a by 2050 in the EU, of which almost 250~Mt/a is sequestered.  The European Innovation Fund~\cite{europeaninnovationfund} supports carbon management technologies to enhance the global competitiveness of European industries.

At the same time, collaborative industry initiatives like the European Hydrogen Backbone~\cite{gasforclimateEuropeanHydrogenBackbone2022} or the Hydrogen Infrastructure Map~\cite{H2InfrastructureMap} showcase the potential of hydrogen as a fuel and energy carrier.
% Some natural gas pipelines have already been repurposed to transport hydrogen~\cite{RohrFreiFuer}.
Business models from companies like Tree Energy Solutions~\cite{TESHydrogenLife2023}, Carbfix~\cite{WeTurnCO2}, and Equinor~\cite{adomaitisEquinorRWEBuild2023} advertise carbon management hubs that provide green hydrogen, methane, and synthetic fuels on the one hand and offer to purchase \carbon{} on the other hand. The Northern Lights project in Norway~\cite{NorthernLightsWhat} is planning a transport and sequestration capacity of 1.5~Mt/a \carbon{} to be operative in 2024, expanding to a targeted scale of 5~Mt/a of sequestration by 2030.
The market potential for carbon capture and sequestration in Europe is underscored by the Capture Map's estimated potential of 1.7 Gt carbon annually from point sources~\cite{ToolsGreenTransition}, accounting for about half of the continent's emissions, as well as estimated sequestration capacity potentials of up to 3~Gt/a~\cite{europeancommissionEuropeanCO2Storage}. However, it is important to note that carbon sequestration technology, unlike carbon capture, is not yet fully mature and its full potential and implications remain to be explored in depth.
To advance the carbon economy, the Clean Air Task Force emphasizes the importance of developing a carbon transport system in Europe alongside a hydrogen network~\cite{lockwoodEuropeanStrategyCarbon}. Carbon pipelines, considered a mature technology, have seen widespread installations in the United States and Canada, primarily to supply \carbon{} to enhanced oil recovery~\cite{righettiSitingCarbonDioxide2017,friedmann2020net}.

Despite the abundance of policies and initiatives, up to this point, it remains unclear how the two transport systems of hydrogen and carbon may complement or even compete with each other. Both can bridge the misalignment of sources and sinks for carbon and hydrogen. A hydrogen network can supply regions with geographically fixed hydrogen demand, such as for steel production, with hydrogen from regions with the best renewable resources as well as enable CU at the site of point-source CC. On the other hand, the carbon transport system allows transporting captured carbon to regions with sequestration potentials or high-quality renewable resources, the latter enabling cost-effective CU.\@

In the literature, the two network approaches and underlying technologies have been discussed in several publications~\cite{bakkenLinearModelsOptimization2008,morbeeOptimisedDeploymentEuropean2012,stewartFeasibilityEuropeanwideIntegrated2014,oeiModelingCarbonCapture2014,elahiMultiperiodLeastCost2014,burandtDecarbonizingChinaEnergy2019,middletonSimCCSOpensourceTool2020,bjerketvedtOptimalDesignCost2020,weiProposedGlobalLayout2021,damoreOptimalDesignEuropean2021,becattiniCarbonDioxideCapture2022,neumannBenefitsHydrogenNetwork2022}. Such techno-economic models, in contrast for example to integrated Assessment Models, can account for the spatial distribution of carbon sources and sinks, which are crucial to provide a holistic view of the energy system and its technological interactions. However, all the techno-economic studies have only dealt with isolated aspects of hydrogen and carbon networks. Neumann et al.~\cite{neumannBenefitsHydrogenNetwork2022} demonstrated the extent to which hydrogen and electricity grid expansion is interchangeable in a climate-neutral European energy system at the expense of higher investments. The underlying, highly-resolved model encompasses the most relevant sectors, however it neglects the infrastructure required to transport carbon across regions.
Morbee et al.~\cite{morbeeOptimisedDeploymentEuropean2012} optimizes the topology and capacity of a \carbon{} network in Europe, but only considers the power sector without co-optimizing renewable deployment. This limited sectoral scope cannot capture important dynamics of carbon management, since it neglects sectors like industry that will need to handle most \carbon{} in the future.
The European Commission's Joint Research Center analyzed the possible topology of a \carbon{} network in 2024~\cite{jrc2024}, but did not include the interaction of  \carbon{} sources and sinks with the rest of the energy system.
Becattini et al.~\cite{becattiniCarbonDioxideCapture2022} present a mixed-integer model to optimize the time-evolution of a \carbon{} transport system in Switzerland, connected to a remote sequestration site in Norway. However, this limited spatial scope fails to consider other sequestration sites and the co-benefits from connecting the \carbon{} network to neighboring countries.
Other high-resolution energy system models do not feature detailed CU technologies and \carbon{} transport. In Pickering et al.~\cite{pickeringDiversityOptionsEliminate2022}, the Euro-Calliope model, for example, features neither carbon nor hydrogen transport, and thus forces generic synthetic fuel production to rely on locale electrolysis and DAC in the same region.

\begin{figure}[h!]
    \includegraphics[width=\linewidth]{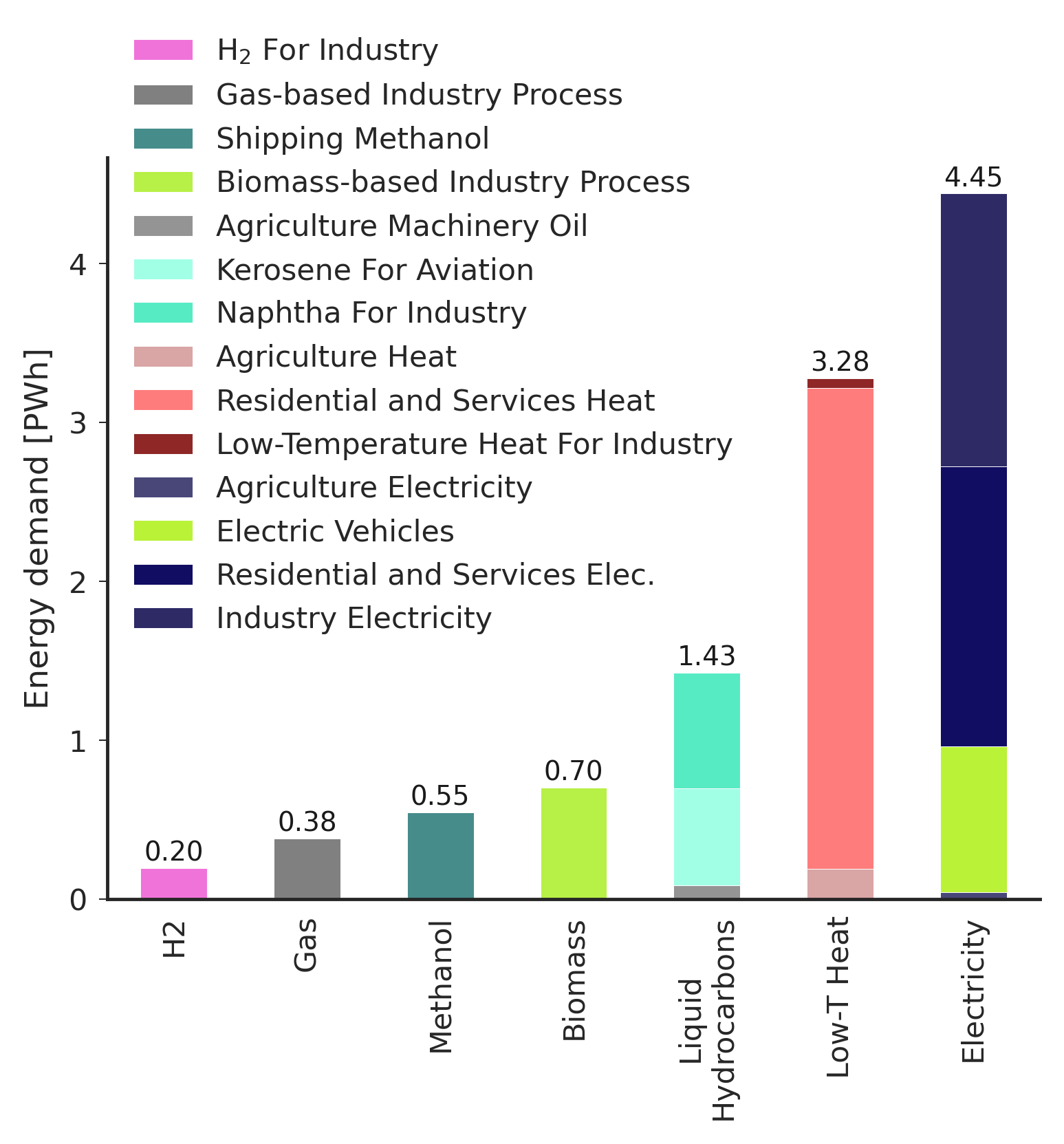}
    \caption{Assumptions on exogenous demand, derived from~\cite{piamanzGeoreferencedIndustrialSites2018,muehlenpfordtTimeSeries2019,mantzosJRCIDEES20152018,NationalEmissionsReported2023,EurostatCompleteEnergyBalance,uwekrienDemandlib2023}. The figure shows total annual energy demands for each energy source, which determine the model's endogenous investments and operation. Endogenous processes can lead to higher total production volumes of some energy carriers, e.g., the demand for methanol requires more hydrogen and carbon as secondary (energy) inputs, which are not considered here. In the model, demands are defined per region and time stamp.}
    % TODO: adjust labels to show that these are exogenous assumptions
    \label{fig:total-demand-bar}
\end{figure}

No study has yet considered the co-optimization and comprehensive assessment of both \carbon{} and \hydrogen{} networks in a fully sector-coupled energy system. Such an assessment is strongly needed to identify realms of competition and synergy between hydrogen and carbon management technologies. In this paper, we present a detailed study of the European energy system for 2050, which includes high geographical resolution and a comprehensive representation of carbon management technologies. The study is conducted using the PyPSA-Eur energy system model~\cite{brownPyPSAEurOpenSectorcoupled2023} which optimizes operations and investments in all relevant energy sectors to supply projected energy demands for 2050 (see Fig.~\ref{fig:total-demand-bar}). It is assumed that Europe is self-sufficient in energy and does not import any fuels. This drives carbonaceous fuel production to be located at sites where both \carbon{} and \hydrogen{} can be provided at low cost.

% TODO: This drives all CCU....
% TODO: The demands should be mentioned properly and introduced

To investigate the competing transport dynamics between \carbon{} and \hydrogen{} networks, we contrast the following four scenarios with different sets of networks present:

\begin{table}[ht!]
    \centering
    \begin{tabular}{c|c|c}
        \toprule
        \textbf{Scenario} & \textbf{\carbon{} Network} & \textbf{\hydrogen{} Network} \\
        \midrule
        Baseline          & --                         & --                           \\
        \carbongrid{}     & \checkmark                 & --                           \\
        \hydrogengrid{}   & --                         & \checkmark                   \\
        Hybrid            & \checkmark                 & \checkmark                   \\
        \bottomrule
    \end{tabular}
    \caption{Set of transportation networks in the developed scenarios.}
    \label{tab:scenarios}
\end{table}

We analyze how the option to deploy carbon and hydrogen networks affects different decarbonization strategies in the scenarios and how investments in none or only one of the transport technologies lead to inefficient infrastructure layouts. Initially, we focus on a net-zero carbon emissions target, limiting carbon sequestration to 200~Mt/a. This constraint ensures that carbon sequestration is reserved for the hardest-to-abate emissions, while avoiding reliance on sequestration where other mitigation options are feasible. We give a detailed description of the different technological and geographical impacts of the transportation systems. Subsequently, our analysis demonstrates the extent to which different deployment strategies remain robust under a tightened target with net \carbon{} removal.

\section{Results}
\label{sec:results}

\begin{figure}[ht!]
    \centering
    \includegraphics[width=\linewidth]{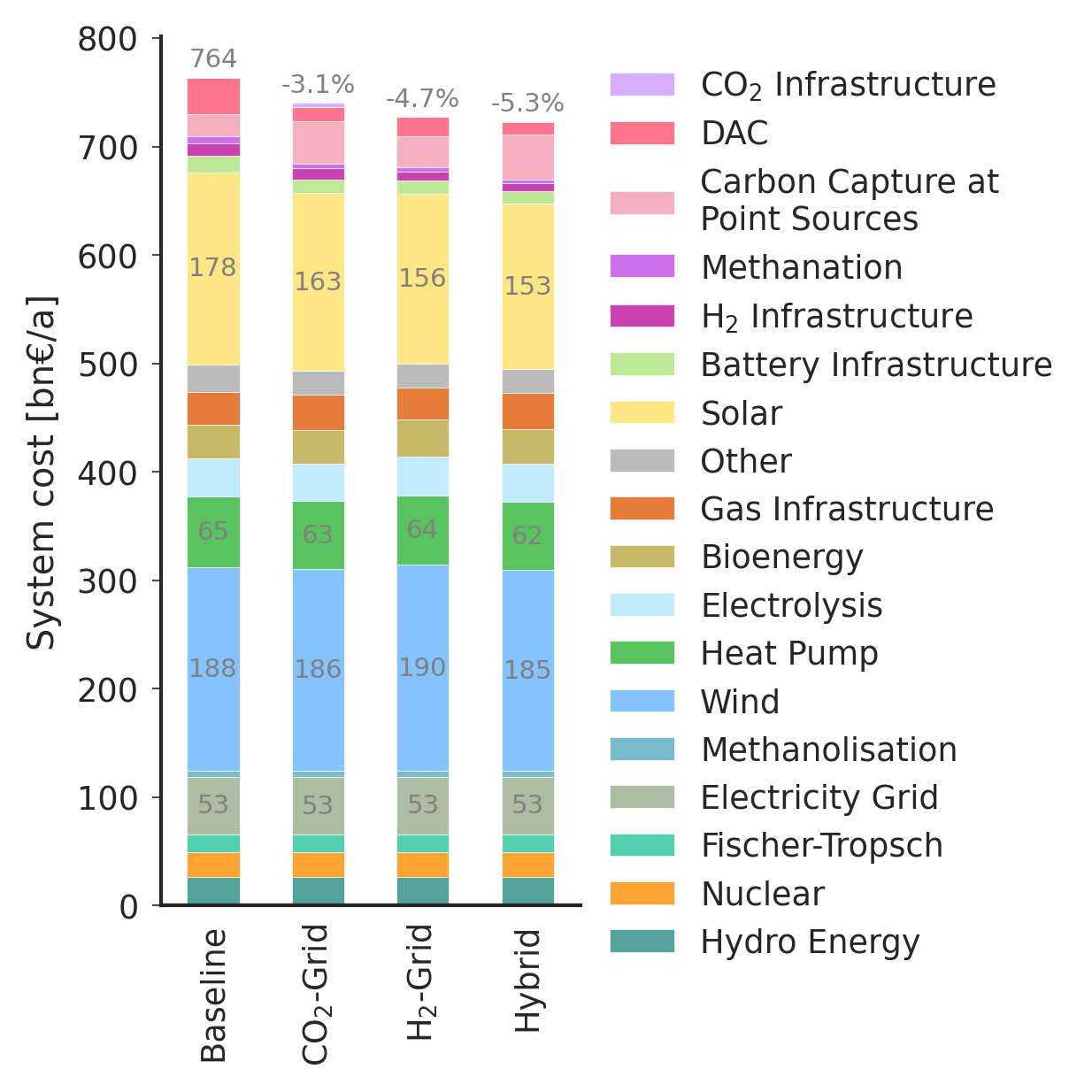}
    \caption[short]{Total annual system cost, subdivided into groups of technologies for the different models of the European energy system with a net-zero emission target. While in the \baselinescenario{}, the model has neither a carbon nor a hydrogen network, it can expand both in the \hybridscenario{}. ``Gas Infrastructure'' combines gas facilities for transport, power and heat production, ``\carbon{} Infrastructure'' and ``\hydrogen{} Infrastructure'' combine transport and storage for each carrier. ``Carbon Capture at Point Sources'' combines all technologies with integrated carbon capture, including the cost of the main facility (e.g., Combined Heat and Power units) and the carbon capture application.}
    \label{fig:cost_bar}
\end{figure}

All scenarios share key characteristics but differ in system costs and technology use (Fig.~\ref{fig:cost_bar}). Costs in the \baselinescenario{} are the highest at 764~bn€/a. Scenarios with carbon and hydrogen networks are more cost-effective, reducing costs by 3.1\% in the \carbongrid{}, 4.7\% in the \hydrogengrid{}, and 5.3\% in the \hybridscenario{}.

% In every model, nearly half the costs are attributable to wind and solar electricity production, and about 10\% to hydro, nuclear, and biomass. In the \baselinemodel, transmission costs (electricity, gas, hydrogen, carbon) are below 4\% (25 bn€) and are dominated by electricity grid expenses. In contrast, the other models have higher combined pipeline costs, peaking at 20\% in the \hybridscenario. Both carbonaceous fuel production and electrolysis account for about 4\% (30 bn€) and 4.7\% (37 bn€) of costs respectively, with heat pump installations contributing around 8\% (64 bn€/a in the \baselinemodel).

\begin{figure}[ht!]
    \centering
    \includegraphics[width=\linewidth]{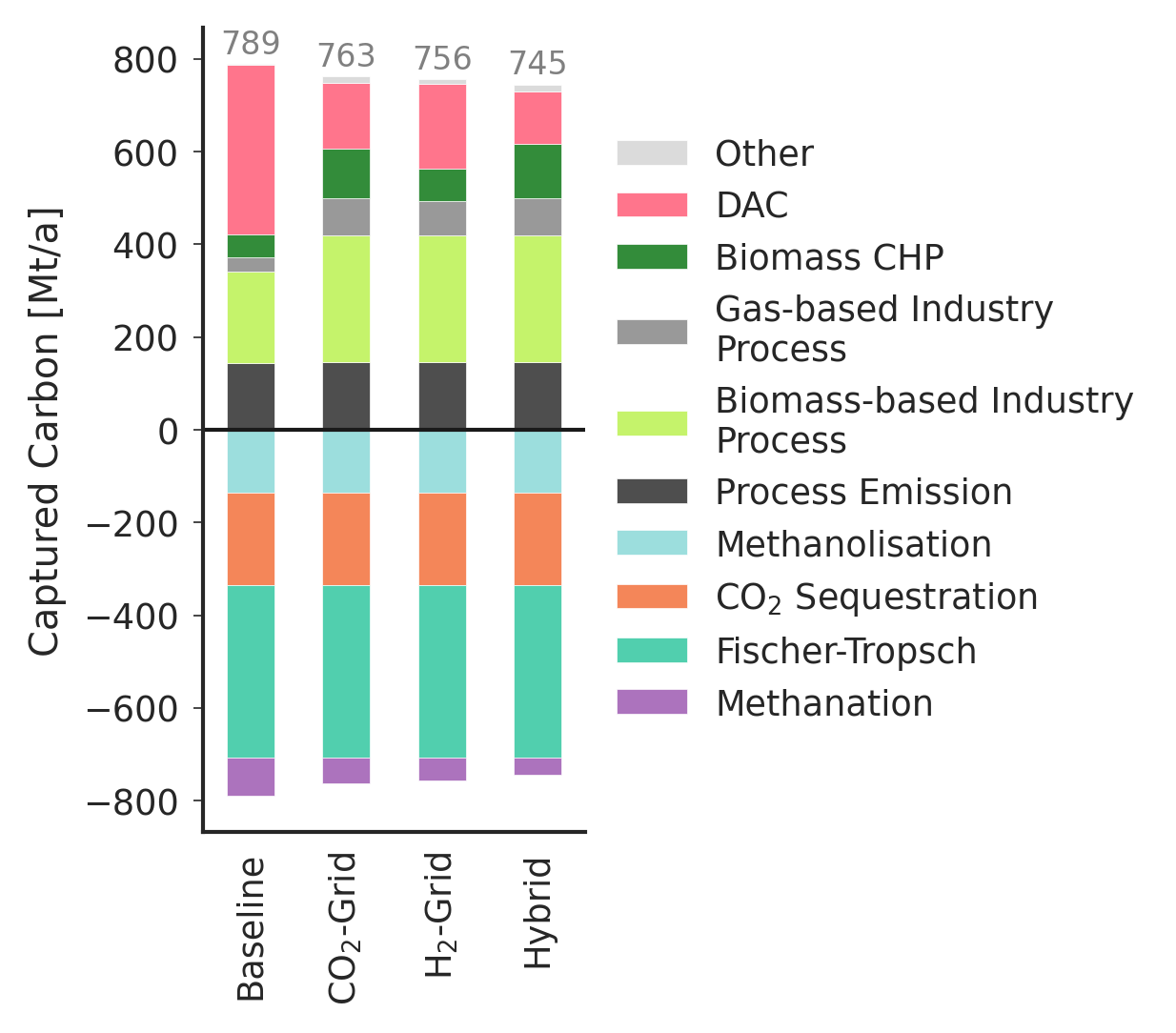}
    \caption{Balance of captured carbon for all scenarios assuming net \carbon{} neutrality. Positive values indicate carbon capture and negative values indicate carbon consumption. By integrating hydrogen and carbon networks, the predominant method for carbon removal shifts from Direct Air Capture (DAC) to bio-energetic processes with capture. At the same time, the reliance on methanation decreases.}
    \label{fig:balance_captured_carbon}
\end{figure}

System cost differences between the scenarios are driven by investments in wind and solar, as well as carbon capture technologies. In scenarios with carbon and hydrogen networks, the model primarily reduces the reliance on costly DAC, from 365~Mt/a carbon capture in the \baselinescenario{} to 113~Mt/a in the \hybridscenario{} (see Fig.~\ref{fig:balance_captured_carbon}). At the same time, carbon capture from biomass combustion, gas-based industry processes and biogas upgrading increase. The latter lowers the need for synthetic methane production.
% Another knock-on effect of less reliance on DAC is its reduced endogenous demand for dedicated renewables as well as heat pumps and biomass CHPs to supply power and heat.

In the \baselinescenario{}, where there is no carbon and hydrogen transport, hydrogen must be produced where it is consumed. For CU, either hydrogen must be produced where carbon is cheap at point sources, or DAC must be used at sites with low-cost hydrogen. Where fossil carbon from point sources cannot be sequestered, it has to be compensated elsewhere by negative emissions from DAC or biomass capture with sequestration. To provide hydrogen in some regions with poor renewable resources the model resorts to SMR with either green or fossil methane transported in the gas grid. All of these options are suboptimal from a cost perspective.

\subsection*{Networks unlock low-cost \hydrogen{} and \carbon{}}

\begin{figure*}[ht!]
    \centering
    \includegraphics[width=\linewidth]{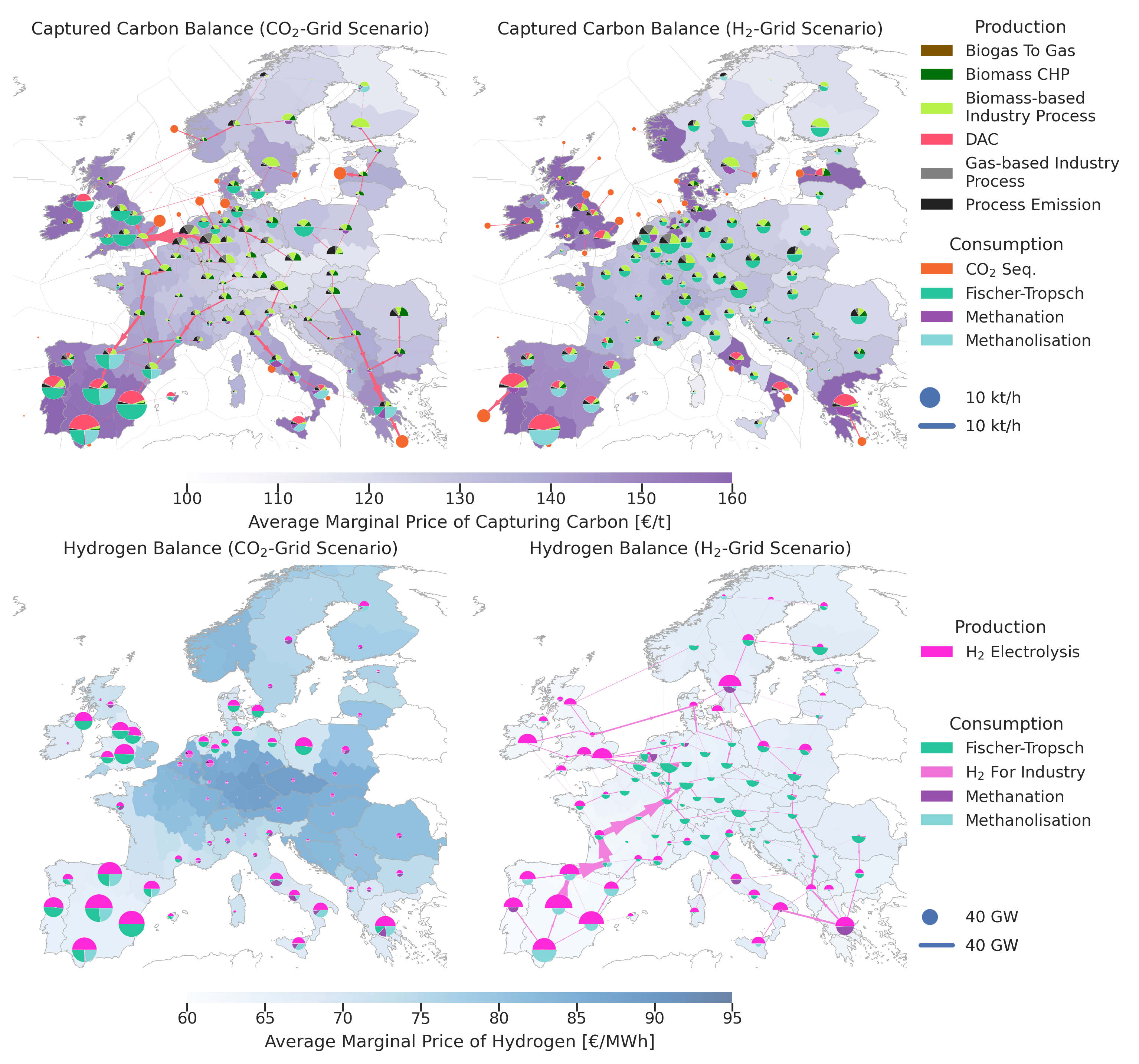}
    \caption{Average production, consumption, flows and prices of the carbon (top line) and hydrogen (bottom line) sectors in the \carbongrid{} (left) and the \hydrogenscenario{} (right). For each region, upper semicircles show the average production per technology, lower semicircles the consumption, and colors the average marginal prices. Lines and arrows show the interregional transportation. Carbon sequestration offshore is drawn in full circles.
    }
    \label{fig:balance_map}
\end{figure*}

Both a carbon network and a hydrogen network can help alleviate these inefficiencies. On the one hand, a carbon network allows carbon from point sources to be transported to low-cost hydrogen as well as to sequestration sites. On the other hand, a hydrogen network allows low-cost hydrogen to be transported from regions with good renewable resources to point sources for CU. This results in fundamentally different network flows in each scenario (see Fig.~\ref{fig:balance_map}).

In the \carbonscenario, there are three main transport purposes of the low-cost carbon captured in Central and Eastern Europe (see Fig.~\ref{fig:balance_map}, top left). A large part of the carbon is transported to the Iberian Peninsula (avg. 9.8~kt/h) and the British Isles (avg. 11.8~kt/h), where it is used along with low-cost electrolytic hydrogen to produce Fischer-Tropsch fuels and methanol. In Italy and Greece, a portion of the carbon (avg. 3.3~kt/h) is directed to methanation facilities where the synthetic methane is fed into the gas grid at key terminals that also receive fossil gas imports. Finally, a third part of captured carbon is transported directly to sites in the North Sea, Baltic Sea and Mediterranean Sea where it is sequestered. Almost none of the captured carbon is used locally at the point source. In addition to electrolysis in regions with good renewable resources to supply CU ($\sim$125~GW$_\text{el}$), the model places smaller amounts of electrolyzers where hydrogen is needed for industrial processes across Central Europe ($\sim$8~GW$_\text{el}$) (see bottom left). Hydrogen prices vary widely across the continent, with low price ``valleys'' at around 60~€/MWh, where the CU is located, and high price ``peaks'' in Central Europe at around 95~€/MWh. Despite the transport system, marginal cost of carbon capture still vary from 105~€/t in Central Europe to 160~€/t at sequestration sites.

In the \hydrogenscenario{}, the main hydrogen transportation route goes from Spain (66~GW) to Central Europe to supply local carbonaceous fuel production with captured carbon from point sources, since \carbon{} can no longer be transported (see Fig~\ref{fig:balance_map}, bottom right). Also, the hydrogen network supplies spatially fixed hydrogen demand for industry in Central Europe (2.5~GW in total) at much lower costs than local electrolysis or SMR. Carbon sequestration takes place at several coastal sites with local DAC facilities. The largest carbon sink is located in Portugal (3.5~kt/h, see Fig~\ref{fig:balance_map}, top right). Marginal costs of carbon capture are low in Central Europe (110-120~€/t) and highest in sequestration regions (160~€/t). Hydrogen prices have flattened out at a relatively low level of €60~/MWh in all regions.

In both scenarios, CU is placed where material inputs, \carbon{} and \hydrogen{}, are provided at minimal cost, i.e., in price valleys of hydrogen and carbon in the \carbongrid{} and \hydrogenscenario{} respectively. In the \carbonscenario{}, CU co-locates with low-cost hydrogen production and uses transported carbon from the inland. In the \hydrogenscenario{}, CU co-locates with carbon point sources and uses transported low-cost hydrogen.

% CU also has an incentive to co-locate in Central Europe with district heating demand, which can use the waste heat from the Fischer-Tropsch process.

% Both scenarios effectively combine low-cost renewable hydrogen production and carbon management for combustion and spatially-fixed industrial processes.

We conclude, there are individual cost-benefits of the transport networks. On the one hand, the carbon network enables the transport of low-cost captured carbon to sequestration sites, reducing the dependence on high-cost DAC near the coast which is present in the \hydrogenscenario{}. On the other hand, the \hydrogengrid{} is able to supply the spatially-fixed hydrogen demands in Central Europe with low-cost hydrogen imports, while the model in the \carbonscenario{} is forced to deploy electrolysis in regions with poor renewable resources. The latter aspect finally leads to lower system costs in the \hydrogenscenario{} than in the \carbonscenario{}.

\subsection*{Hybrid configuration combines advantages}\label{subsec:Hybrid}

\begin{figure*}[ht!]
    \centering
    \begin{subfigure}{.5\textwidth}
        \centering
        \includegraphics[width=\linewidth]{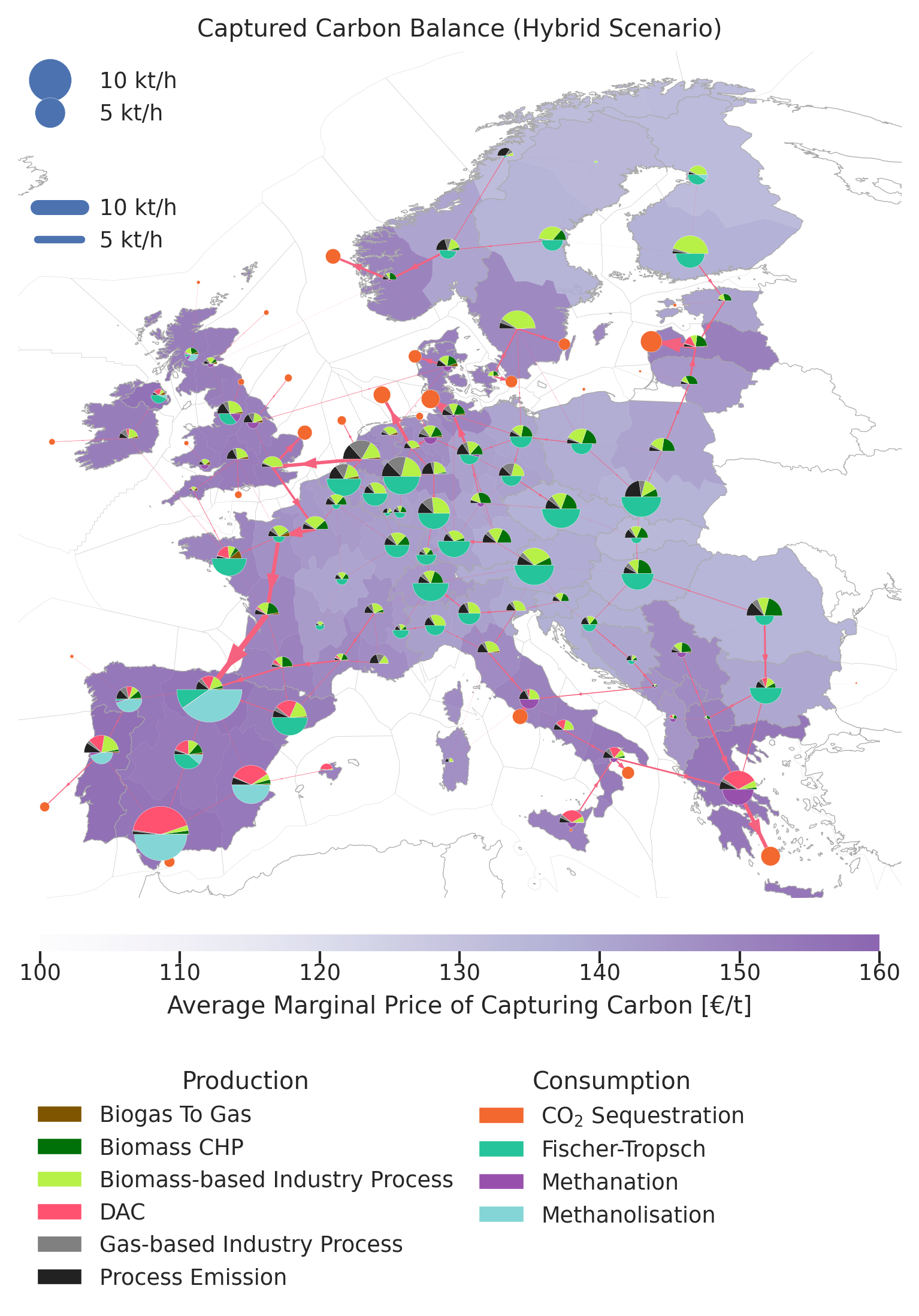}
        \label{fig:balance_map_carbon_full}
    \end{subfigure}%
    \begin{subfigure}{.5\textwidth}
        \centering
        \includegraphics[width=\linewidth]{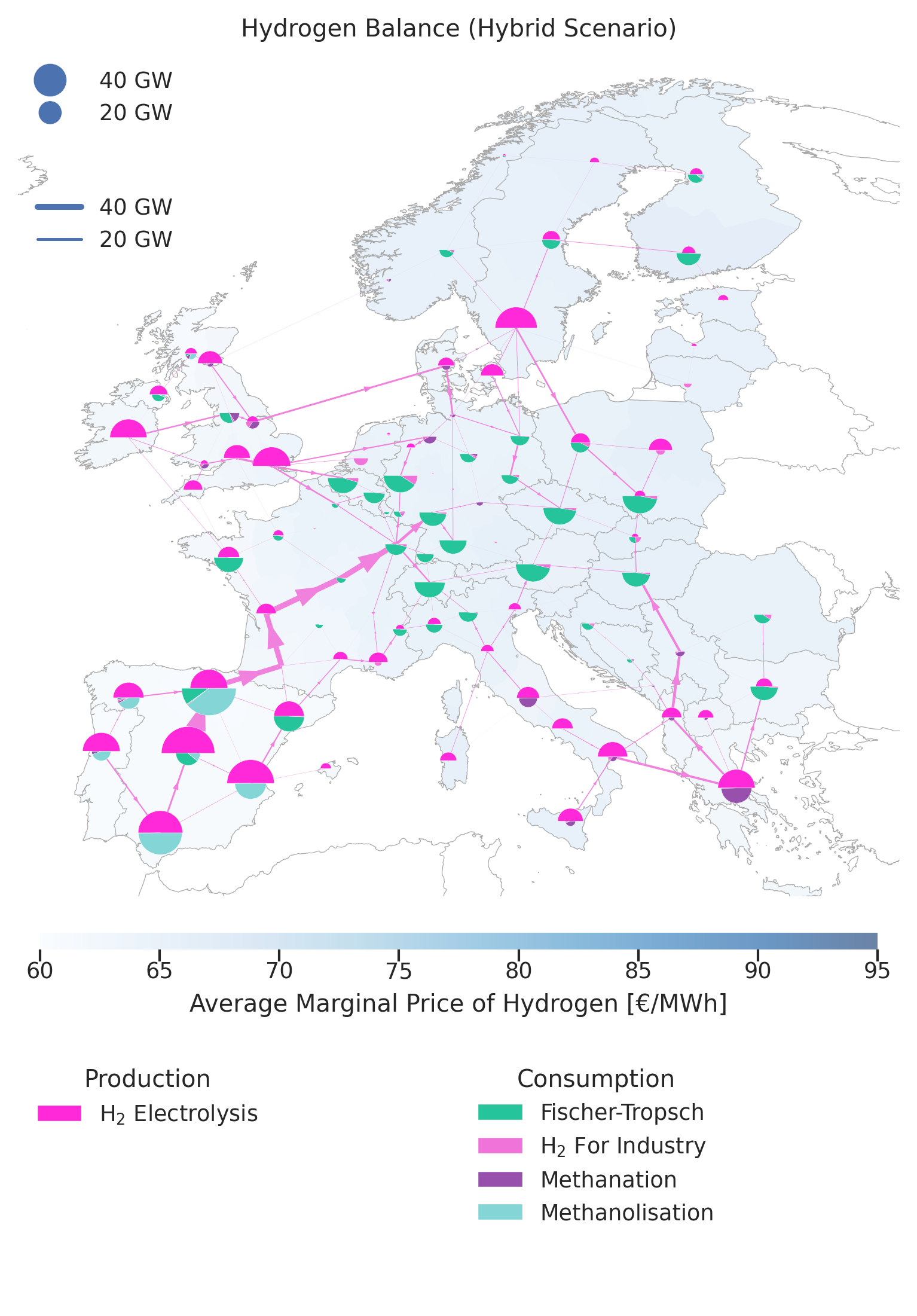}
        \label{fig:balance_map_hydrogen_full}
    \end{subfigure}
    \caption{Average operation, flows and prices of the carbon (left) and hydrogen (right) sectors in the \hybridscenario{} assuming net \carbon{} neutrality. For each region, upper semicircles show the average production per technology, lower semicircles the consumption, and colors the average marginal prices. Carbon Sequestration offshore is drawn in full circles. Lines and arrows show interregional transportation volume. Carbon from point-source in the inland either supplies local CU with imported hydrogen or facilitates sequestration in nearby offshore regions.
        % Carbon network looks the same as in~\cite{morbeeOptimisedDeploymentEuropean2012}: two backbones, one in northern Europe other in south east.
    }
    \label{fig:balance_map_full}
\end{figure*}

The \hybridscenario{} combines the advantages of hydrogen and carbon networks, resulting in the highest system cost reduction of 5.3\% (41~bn€/a) compared to the \baselinescenario{}. Again, the model transports low-cost hydrogen to the point sources in Central Europe to supply CU, but also transports low-cost carbon from point sources close to the shore to sequestration sites (see Fig.~\ref{fig:balance_map_full}). This provides low-cost hydrogen across the regions and reduces reliance on more expensive DAC.

For the carbon network, the model employs short and direct pipeline routes to transport carbon from inland point sources to nearby sequestration sites, for example in Northern Germany, Italy and Greece (see Fig.~\ref{fig:balance_map_full}, left). In addition, pipelines transport rather small amounts of carbon from spatially fixed point sources to CU facilities. A notable outlier is the route from the Netherlands via the UK and France to large methanolization and Fischer-Tropsch plants in northern Spain.
The hydrogen routes are similar to those in the \hydrogenscenario{} with transports from Western regions (the Iberian Peninsula and the British Isles) to distributed CU production sites across the continent (see Fig.~\ref{fig:balance_map_full}, right).
While the topology of the hydrogen network (1.2~bn€/a) roughly corresponds to that of the \hydrogenscenario{} and is primarily driven by CU, the carbon network (0.6~bn€/a) plays a secondary role and is driven by CS (compare Fig.~\ref{fig:cost_bar_transmission}).

\begin{figure}[ht]
    \centering
    \includegraphics[width=\linewidth]{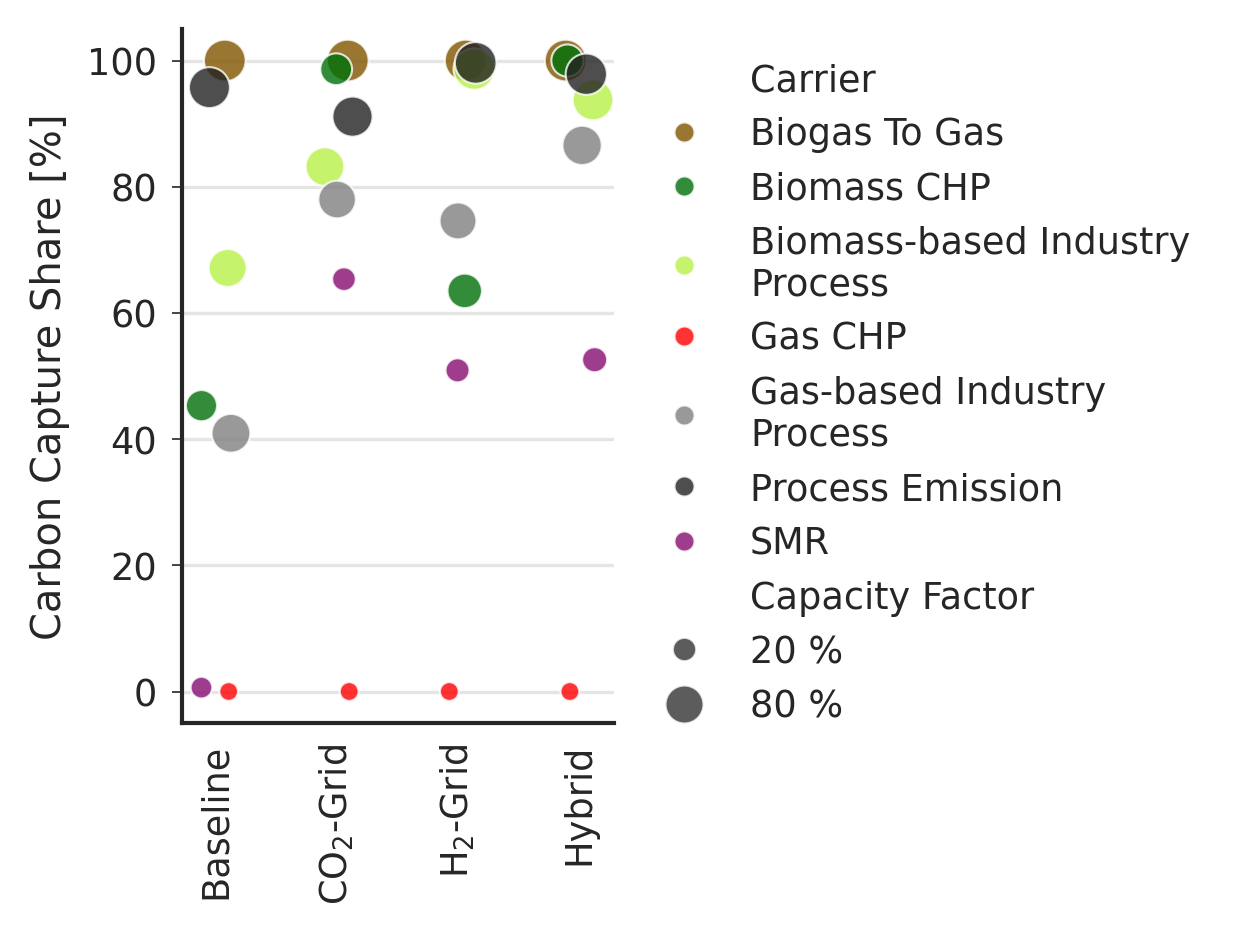}
    \caption{Proportion of plants with integrated carbon capture for scenarios. The size of the dots corresponds to the average capacity factor of the respective technology. While peak load technologies such as gas-fired combined heat and power (CHP) plants, which are only in operation a few weeks a year, are not expanded for carbon capture, point sources with base load characteristics such as biomass CHP and process emissions from industry are fully (partially) developed as soon as the transport of carbon (hydrogen) is permitted.}
    % A CO2 network unlocks BECSS potentials, most biomass emissions are captured at point sources and transported to sequestration sites. cite~\cite{rosaAssessmentCarbonDioxide2021}
    \label{fig:captureshare_line}
\end{figure}%

In this setup, the networks enable all carbon-emitting technologies that operate at a high average capacity factor, i.e., above 70\%, to capture most of their carbon (see Fig.~\ref{fig:captureshare_line}). More than 90\% of emissions are captured for most technologies. The exceptions are SMR and gas-fired CHP plants, which operate only a few weeks annually, making carbon capture units economically unfeasible for these applications.
Due to the better integration of point sources, only a few DAC facilities are installed, such as those used for methanolization in Spain. This, in turn, leads to reduced power consumption and less wind and solar power deployment than in the other scenarios.

\subsection*{Network layouts are robust against tighter emission targets}
\begin{figure}[htb!]
    \centering
    \includegraphics[width=0.9\linewidth]{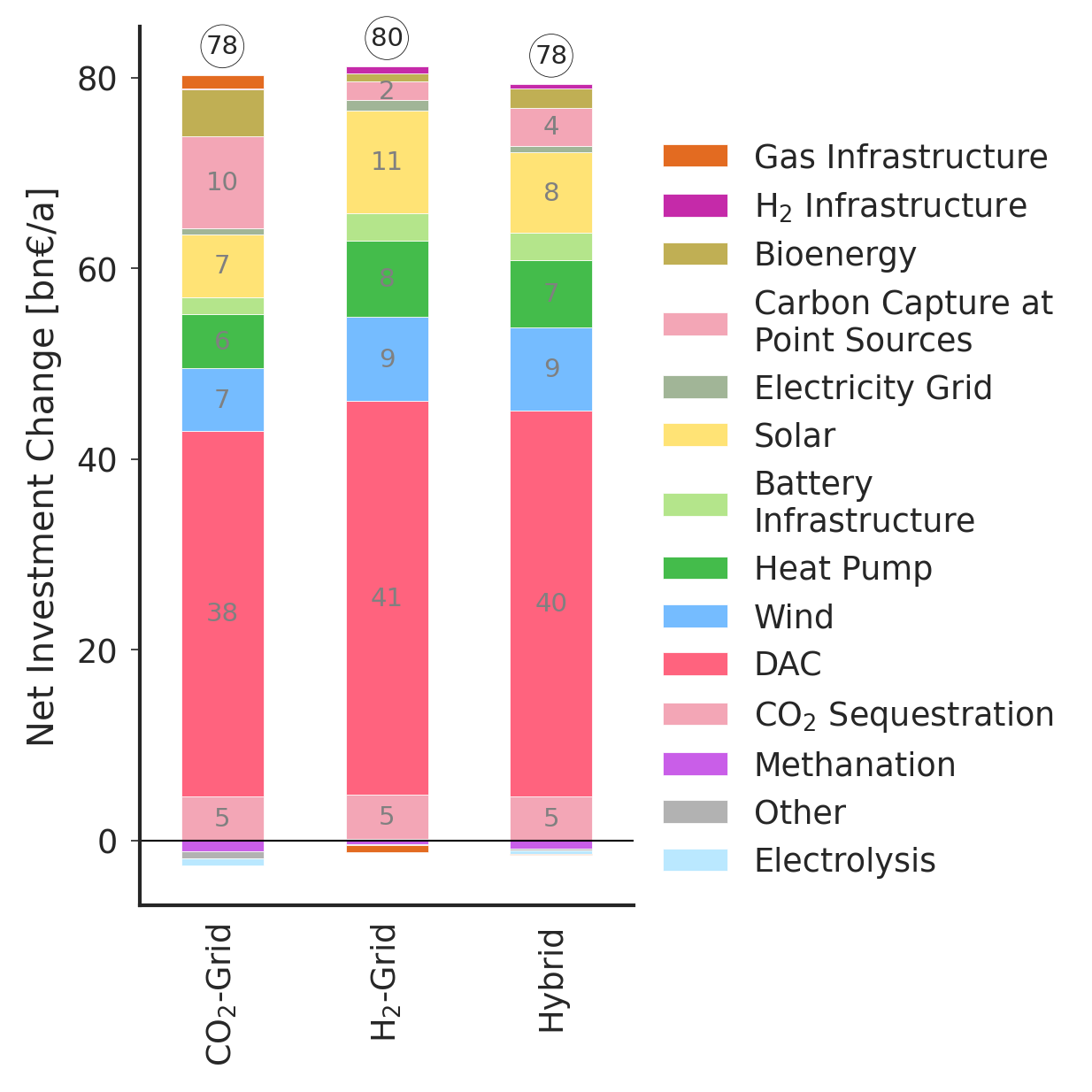}
    \caption[short]{Net change in system cost when tightening the \carbon{} emission target from net-zero to net-negative 10\% of 1990s emissions. Each cost bar is split into contributions of the same technology groups as used in Fig.~\ref{fig:cost_bar}, except for carbon sequestration costs, which are now displayed separately. For all models, Direct Air Capture (DAC) contributes the most to additional \carbon{} removal, requiring further solar, wind and heat pump capacities for electricity and heat input.}
    \label{fig:net-negative_cost_bar}
\end{figure}

In a further set of scenarios, we apply a stricter \carbon{} emissions target and force the model to remove 460~Mt of \carbon{} in one year (equivalent to 10\% of 1990 \carbon{} emissions) while sequestering a maximum of 660~Mt. There are additional system costs of 78--82~bn€/a in all scenarios, of which the largest proportion is the investment in DAC and associated solar and wind power and heat supply (see Fig.~\ref{fig:net-negative_cost_bar}). In the \hybridscenario{}, this share is 78\%, while new carbon capture at point sources accounts for only 4\% of incremental system costs. This shows that point source carbon capture is already almost exhausted in the net-zero case. The additional sequestration is mainly provided by large DAC facilities near the coast. To a much lesser extent, carbon captured from inland bioenergy sources, which would provide CU under a net-zero assumption, is instead transported and sequestered (see Fig.~\ref{fig:balance_maps_full_nn}). Some of the CU located in Central Europe in net-zero scenarios is instead diverted to Spain, where it uses carbon from new DAC facilities.

These changes have little effect on the layout of the \carbon{} and \hydrogen{} networks (see Fig.~\ref{fig:cost_map_difference_full_nn} and Fig.~\ref{fig:balance_maps_full_nn}).
The slight shift in CU deployment results in less hydrogen pipeline construction in some parts of the system. Nevertheless, 80\% of the hydrogen topology built under the net-zero emissions assumption is also built under the net-negative target. The \carbon{} network partially extends routes from Central and Eastern Europe to the coasts. The relative benefits of \carbon{} transport and sequestration with tighter emissions targets are evident when compared with the other scenarios. The carbon network in the net-negative \carbonscenario{} is larger than in the net-zero \carbonscenario{} and transports more carbon from point sources to sequestration sites than to CU facilities. In the \hydrogenscenario{}, where all carbon for CU and CS must come from DAC, the hydrogen transport topology remains almost identical to the net-zero scenarios (for detailed cost contributions see sections~\ref{sec:cost_comparison} and~\ref{sec:operation_nn} in the Appendix).

Overall, this leads to the net cost increases of 78~bn€/a in the \carbonscenario{}, by 80~bn€/a in the \hydrogenscenario{}, and by 78~bn€/a in the \hybridscenario{}. The resulting cost-benefit of the \hybridscenario{} in comparison to the other scenarios is 6~bn€/a.

\section{Discussion}
\label{sec:conclusion}
Our study assesses the roles of hydrogen (\hydrogen{}) and carbon dioxide (\carbon{}) networks in Europe's future energy system under net carbon neutrality and net-removal targets. We show that both networks strongly impact the optimal deployment of carbon capture, carbon utilization and sequestration. A \carbon{} network reduces costs by transporting low-cost carbon from distributed point sources to sequestration sites and sites with low-cost hydrogen production to produce carbonaceous fuels. A hydrogen network provides greater cost savings as low-cost hydrogen can be transported to regions with high demand, allowing carbon utilization at point sources. A combination of both networks emphasizes the strengths of each infrastructure. In the hybrid configuration, the \hydrogen{} network supplies low-cost hydrogen to spatially fixed hydrogen demand and carbon utilization at point sources across regions. At the same time, the \carbon{} network transports low-cost carbon captured from point sources near the coast to sequestration sites, reducing the reliance on direct air capture. This results not only in the lowest system cost but also in a system robust to tightening the target to net-negative emissions.

Limitations of the study, such as the neglect of synthetic fuel imports and assumptions on biomass potentials, are discussed in the Appendix in Section~\ref{sec:limitations}.
The Appendix in Section~\ref{sec:sensitivities} provides an analysis of the model's sensitivities. Notably, relaxing the limit from 200 to 800~Mt/a reduces system costs by 9.1\% in the hybrid scenario. This reduction is achieved by replacing the production of synthetic e-fuels with fossil fuels that are either captured and sequestered, or left unabated and compensated by carbon dioxide removal. Investments in renewable power assets and hydrogen infrastructure which serve the synthetic fuel production for the aviation, shipping and industry sectors drop by roughly a third. A similar effect on synthetic fuel production in Europe would be expected if synthetic fuels were instead imported from outside Europe. Smaller effects can be seen from reducing capital costs for \carbon{} pipelines and costs of electrolyzers.

For policymakers these results show the need for coordinated planning across sectors like carbon, hydrogen and synthetic fuels since they are strongly interacting, as well as across borders because of the localized nature of good renewable resources and sequestration potentials. Deploying multiple networks offers some robustness should unforeseen problems arise with one of them, and our results show that the system can even cope with neither a \carbon{} nor \hydrogen{} network for a cost penalty.  Many of the technologies have not yet been deployed at scale before and will need support for financing, regulation and gaining the acceptance of local populations.

In summary, our study highlights the benefits of strategic planning and integration of both \carbon{} and \hydrogen{} networks in the European energy landscape. The flexibility and cost-efficiency offered by these networks can make a significant contribution to achieving the EU's climate targets, in particular by reducing dependence on technologies that are still in the early market phase, such as Direct Air Capture. Leveraging the synergies between \hydrogen{} and \carbon{} transportation and management, Europe can effectively reduce system costs while ensuring a robust and sustainable energy system.

\section*{Methods}
\label{sec:methodology}

The study is conducted based on the open-source, capacity expansion model PyPSA-Eur~\cite{horschPyPSAEurOpenOptimisation2018,brownSynergiesSectorCoupling2018,PyPSAEurSecSectorCoupledOpen2023}.
The model uses the open-source software tools PyPSA~\cite{brownPyPSAPythonPower2018} and Linopy~\cite{hofmannLinopyLinearOptimization2023}
for modeling and cost optimization.
The model optimizes the design and operation of the European energy system, encompassing the power, heat, industry, agriculture, and transport sectors, including international aviation and shipping.
All technology cost assumptions are taken for the year 2040 and sourced from an open-source database~\cite{zeyenPyPSATechnologydataTechnology2023}. The database applies a discount rate of 7\% for all major technologies, accounting for the weighted average costs of capital (WACC) of investments.
Many of the assumptions contained therein are based on the technology catalogs published by the Danish Energy Agency~\cite{danishenergyagencyTechnologyDataGeneration2019,thedanishenergyagencyTechnologyDataCarbon2023}.
Endogenous model results include expansion of renewable energy sources, storage technologies, Power-To-X conversion and transmission capacities, heating technologies, peaking power plants, and the deployment of gray, blue or green hydrogen production facilitates, among others.
The model considers various energy carriers like electricity, heat, hydrogen, methane, methanol, liquid hydrocarbons, and biomass, together with the corresponding conversion technologies.

% more detail on heating technologies?
% with transport costs considered? - how high are they?

%
In our configuration, the model's time horizon spans one year, with a 3-hourly temporal resolution and a spatial resolution of 90 regions. Each of the regions consists of a complex subsystem with technologies for supplying, converting, storing and transporting energy. Exogenous assumptions on energy demand and non-abatable emissions are taken from various sources~\cite{piamanzGeoreferencedIndustrialSites2018,muehlenpfordtTimeSeries2019,mantzosJRCIDEES20152018,NationalEmissionsReported2023,EurostatCompleteEnergyBalance,uwekrienDemandlib2023} (see Fig.~\ref{fig:total-demand-bar}). The energy demand for electricity, transport, biomass, heat and gas is defined per region and time-step.
Land transport demand is exogenously assumed to be fully electrified, including heavy-duty vehicles.
Demands for kerosene for aviation, methanol for shipping, and naphtha for industry are not spatially resolved and assumed to be constant throughout all time steps.
Heat demand is regionally subdivided into shares of urban, rural and industrial sites.
The system exogenously produces 633~Mt/a \carbon{} from industry, aviation, shipping and agriculture, 153~Mt of which are fossil-based process emissions.
Locations for industrial clusters are taken from~\cite{hotmaps_industrial_db}. Energy demand for industries is calculated from~\cite{mantzosJRCIDEES20152018}.

% Low-carbon electricity potentials

Low-carbon electricity is provided by wind, solar, biomass, hydroelectricity and nuclear power plants. Hydroelectric and nuclear plants cannot be extended beyond their currently installed capacities. Weather-dependent power potentials for solar, wind and hydro-electricity are calculated from the reanalysis and satellite data sets, ERA5 and SARAH-2,~\cite{hersbachERA5GlobalReanalysis2020,pfeifrothSurfaceRadiationData2017} per region and time-stamp, using the open-source tool Atlite~\cite{hofmannAtliteLightweightPython2021}.
Solar and wind power can be expanded in alignment with land-use restrictions calculated, taking into account land usage classes and natural protection areas~\cite{eeaCorineLandCover2012,eeaNatura2000Data2016}. We restrict the total volume of power transmission expansion to 20\% of its current capacity, acknowledging the challenges in inaugurating new transmission projects.
For the use of biomass, we consider only residual biomass products and no energy crops. We limit regional biomass use to the medium-level potentials from the JRC-ENSPRESO database~\cite{enspreso_database,instituteforenergyandtransportjointresearchcentreJRCEUTIMESModelBioenergy2015}. Interregional biomass transport is permitted, with transport costs of $\sim$0.1~€/MWh/km considered.

% energy transport

The topology and capacities of the electricity transmission system are taken from the ENTSO-E transparency map~\cite{wiegmansGridkitExtractEntsoE2016} and selected Ten Year Network Development Plan (TYNDP) projects. The power flow is based on the linearized power flow approximation assuming a linear transport efficiency of 95.75\% per 1000~km accounting for resistive losses on the power lines. %TODO find reference
The topology of the \hydrogen{}, \carbon{} and gas network is identical to the topology of the electricity network, connecting all neighboring regions.
For hydrogen pipelines an average loss of 1.2\% per 1000~km, aligning with industry estimates that account for hydrogen's high permeability and small molecular size, which can lead to higher losses compared to other gases. The electrical demand for compression stations is assumed to be 1.8\% per 1000~km of the transported energy.
For \carbon{} pipelines an average loss rate of 0.8\% per 1000~km is reflecting a generally low leakage rate observed in supercritical or dense phase \carbon{} transport~\cite{liuExperimentalStudyLeakage2023,vitaliRisksSafetyCO22021}. The compression energy needed is set to 250~kWh per 1000~km and tonne \carbon{} to maintain a supercritical state for efficient transport. For natural gas pipelines, the model assumes a 1\% loss rate per 1000~km, accounting for typical leakage and operational losses in well-maintained natural gas infrastructure as reported in~\cite{NaturalGasTransmission2021}. The electrical demand for compression is set at 1\% per 1000~km of the transported energy, which falls within the common range for natural gas pipeline operations~\cite{mcvayreneeMethaneEmissionsGas2023}.
Liquid fuels like oil, methanol and Fischer-Tropsch (FT) fuels are not spatially resolved, since transport costs per unit of energy are negligible due to their high energy density. Throughout this study, we focus on \carbon{} and \hydrogen{} networks because of their high relevance to infrastructure investments and thus public policy decisions. The electricity network and gas network are both included in the model with full geographical detail, but are not further analyzed. Electricity networks are already in place, and we restrict further extension to 25\% due to concerns about public acceptance. Gas networks also already exist, and will most likely experience decreased use in the future, removing any bottlenecks or constraints on the optimal system buildout or operation.
If enabled, hydrogen can be transported via pipelines between regions which can be expanded without limits, considering costs for pipeline segments and compressors. Pipeline flows are modeled using net transfer capacities and without flow dynamics, pressure valves, or energy demand for compression. No retrofitting of gas pipelines is considered.
% gas network data from SciGrid gas

% Modelling of hydrogen supply chain pathways

In our model, we consider green, blue and gray hydrogen from electrolysis and steam methane reforming (SMR), the latter of which may be equipped with CC technology. The geographical distribution of underground hydrogen storage potentials in salt caverns is derived from Caglayan et al.~\cite{caglayanTechnicalPotentialSalt2020}. Re-electrification of hydrogen is possible via fuel cells.

% carbon utilization

We argue that it is crucial for proper modeling to consider CC, CU and CS as separate technologies, rather than grouping them under the often-used but misleading umbrella term ``carbon capture, utilization and storage'' (CCUS), to allow for independent optimization and adequate representation of each function in the carbon management system.
Our model features three drop-in fuel production technologies for CU: methanation, methanolization, and Fischer-Tropsch synthesis.
The processed fuels are not limited in their total quantity of use or production.
Methane is transported through the gas network, while methanol and FT fuel can be transported between regions without additional costs or capacity constraints.
Synthetic methane substitutes natural gas or upgraded biogas, serving combined heat and power plants, residential heating gas boilers, or industrial process heat.
Synthetic methanol decarbonizes marine industry fuel demands, and FT fuels replace fossil oil for naphtha production for high-value chemicals, aviation kerosene, or agricultural machinery oil.

% carbon supply
To supply carbon needed for CS and CU, the system can choose to deploy carbon capture (CC) technologies at various point sources of fossil or biogenic origin (see below), or through DAC facilities.
The concept of a merit order for capturing carbon plays a pivotal role in optimizing the deployment of carbon capture technologies based on their economic feasibility.
This merit order ranks all CC technologies according to their relative costs to capture an additional marginal ton. The cost value for a technology depends on the exact system configuration and operation and may vary by scenario and even by region since many technologies produce other (primary) outputs such as electricity and heat. The costs to capture an additional marginal ton of carbon may also be different from the average cost of capture.
At the lower end of the cost spectrum, CC technologies applied to process emissions, such as those from cement, offer a cost-effective starting point.
Following this, biomass combined heat and power (CHP) systems provide the dual benefit of energy production and carbon capture.
Moving up the scale, gas used in industrial applications and biomass employed in industrial processes represent more costly yet viable options to capture carbon.
DAC, an emerging technology capable of extracting \carbon{} directly from the atmosphere, stands at a high-cost level due to its current high capital costs and energy demand.
Finally, based on our model results, biogas upgrading, a process that refines biogas to natural gas quality, incurs the highest costs in the merit order per marginal ton of captured \carbon{}.
Biogas input is a high-cost fuel that based on endogenous modeling decisions is not used in large quantities.
To the extent that biogas-to-gas facilities are used by the model to supply additional (carbon-neutral) gas as fuel, adding CC infrastructure incurs low costs (and thus ranges on the left-hand side of the merit order curve).
However, the price of capturing additional \carbon{} from biogas-to-gas upgrading is high because a substantial amount of the biogas fuel costs factors into the marginal cost of captured \carbon{}.
This merit order framework is crucial for strategically deploying necessary carbon capture solutions while balancing economic considerations.
If we consider the spatial aspect of distributed carbon capture potentials, the \carbon{} network can exploit comparative cost advantages between \carbon{} bidding zones to utilize the lowest-cost CC facilities across the continent.
We assume a capture rate of 90\% for CC on process emissions, SMR, biogas-to-gas, as well as gas and biomass used in industry, and 95\% for CC on biomass and gas CHPs.

% carbon storage and sequestration

\begin{figure}[h!]
    \centering
    \includegraphics[width=\linewidth]{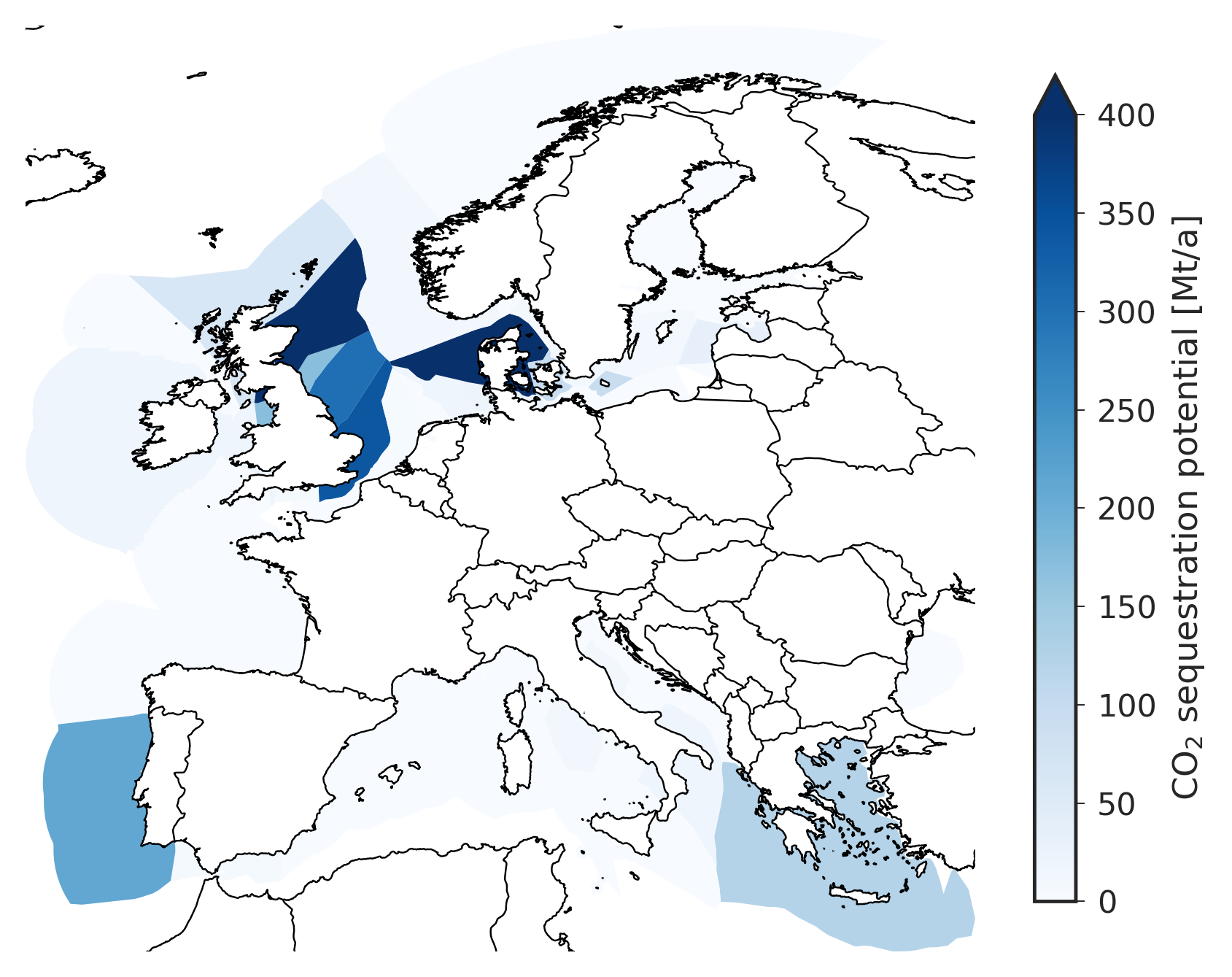}
    \caption{Maximal sequestration potential per offshore region used as input for all models. Note that in most model runs the regional sequestration potentials are not exploited due the the global limit on sequestration.}
    \label{fig:sequestration_map}
\end{figure}

To store carbon, we differentiate between short-term storage in steel tanks without permanent containment and long-term, irreversible sequestration in underground sequestration sites such as porous rock formations or depleted gas reservoirs.
Costs for both options are included in the model.
For carbon sequestration, we only consider offshore sites as potential sinks (see Fig.~\ref{fig:sequestration_map}).
We make this choice because offshore sequestration sites typically have a larger capacity compared to onshore sequestration sites in saline aquifers and due to concerns over public safety for infrastructure near populated areas.
Our estimates for carbon sequestration potentials are restrictive, limiting the annual sequestration potential to 25~Mt per region and constraining it further such that it provides enough volume to allow for 25 years of sequestration.
Furthermore, we decided to limit the total amount of sequestered \carbon{} to 200~Mt/a for our net-zero \carbon{} emission target and 660~Mt/a for our net-negative emission target.
This constraint is deliberately implemented to avoid over-reliance on carbon capture and sequestration as a backstop technology, which could otherwise offset fossil emissions instead of prioritizing the reduction of fossil fuel use where technologically feasible. By capping the sequestration at 200 Mt/a, we aim to promote a balanced energy transition, where sequestration is used to offset hard-to-abate fossil process emissions (e.g., calcination of limestone in cement manufacturing) but does not become a substitute for more sustainable mitigation measures. The implications of this choice are discussed in the Appendix in Section~\ref{sec:higher_sequestration}.

\subsubsection*{Data availability}

All input, processed and resulting data is available or can be generated via the workflow \href{https://github.com/FabianHofmann/carbon-networks}{github.com/FabianHofmann/carbon-networks} and the underlying workflow \href{https://github.com/PyPSA/pypsa-eur}{github.com/pypsa/PyPSA-Eur}. Data on the optimized energy system models analyzed throughout the study was made publicly available under \href{https://doi.org/10.6084/m9.figshare.24923028}{10.6084/m9.figshare.24923028}.

\subsubsection*{Code availability}

The code to reproduce all figures and numbers is available at \href{https://github.com/FabianHofmann/carbon-networks}{github.com/FabianHofmann/carbon-networks}.

\subsubsection*{Acknowledgements}

Fabian Hofmann and Christoph Tries are funded by the Breakthrough Energy Initiative (\href{https://breakthroughenergy.org/}{breakthroughenergy.org}).

\printbibliography

@article{adomaitisEquinorRWEBuild2023,
  title = {Equinor and {{RWE}} to Build Hydrogen Supply Chain for {{German}} Power Plants},
  author = {Adomaitis, Nerijus and Steitz, Christoph},
  year = {2023},
  month = jan,
  journal = {Reuters},
  url = {https://www.reuters.com/business/energy/equinor-rwe-build-hydrogen-supply-chain-german-power-plants-2023-01-05/},
  urldate = {2023-02-01},
  chapter = {Energy},
  langid = {english}
}

@article{bakkenLinearModelsOptimization2008,
  title = {Linear {{Models}} for {{Optimization}} of {{Infrastructure}} for {{CO2 Capture}} and {{Storage}}},
  author = {Bakken, Bjorn H. and {von Streng Velken}, Ingrid},
  year = {2008},
  month = sep,
  journal = {IEEE Transactions on Energy Conversion},
  volume = {23},
  number = {3},
  pages = {824--833},
  issn = {1558-0059},
  doi = {10.1109/TEC.2008.921474}
}

@article{becattiniCarbonDioxideCapture2022,
  title = {Carbon Dioxide Capture, Transport and Storage Supply Chains: {{Optimal}} Economic and Environmental Performance of Infrastructure Rollout},
  shorttitle = {Carbon Dioxide Capture, Transport and Storage Supply Chains},
  author = {Becattini, Viola and Gabrielli, Paolo and Antonini, Cristina and Campos, Jordi and Acquilino, Alberto and Sansavini, Giovanni and Mazzotti, Marco},
  year = {2022},
  month = jun,
  journal = {International Journal of Greenhouse Gas Control},
  volume = {117},
  pages = {103635},
  issn = {1750-5836},
  doi = {10.1016/j.ijggc.2022.103635},
  url = {https://www.sciencedirect.com/science/article/pii/S1750583622000548},
  urldate = {2022-11-29},
  langid = {english}
}

@article{bjerketvedtOptimalDesignCost2020,
  title = {Optimal Design and Cost of Ship-Based {{CO2}} Transport under Uncertainties and Fluctuations},
  author = {Bjerketvedt, Vegard Skonseng and Tomasgard, Asgeir and Roussanaly, Simon},
  year = {2020},
  month = dec,
  journal = {International Journal of Greenhouse Gas Control},
  volume = {103},
  pages = {103190},
  issn = {1750-5836},
  doi = {10.1016/j.ijggc.2020.103190},
  url = {https://www.sciencedirect.com/science/article/pii/S1750583620306150},
  urldate = {2023-03-13},
  langid = {english}
}

@misc{brownPyPSAEurOpenSectorcoupled2023,
  title = {{{PyPSA-Eur}}: {{An}} Open Sector-Coupled Optimisation Model of the {{European}} Energy System},
  shorttitle = {{{PyPSA-Eur}}},
  author = {Brown, Tom and Victoria, Marta and Zeyen, Elisabeth and Hofmann, Fabian and Neumann, Fabian and Frysztacki, Martha and Hampp, Johannes and Schlachtberger, David and H{\"o}rsch, Jonas},
  year = {2023},
  month = jul,
  doi = {10.5281/ZENODO.8189026},
  url = {https://zenodo.org/record/8189026},
  urldate = {2023-12-21},
  copyright = {MIT License, Open Access},
  howpublished = {Zenodo}
}

@article{brownPyPSAPythonPower2018,
  ids = {brown_pypsa:_2018,pypsa},
  title = {{{PyPSA}}: {{Python}} for {{Power System Analysis}}},
  shorttitle = {{{PyPSA}}},
  author = {Brown, Tom and H{\"o}rsch, Jonas and Schlachtberger, David},
  year = {2018},
  month = jan,
  journal = {Journal of Open Research Software},
  volume = {6},
  eprint = {1707.09913},
  pages = {4},
  issn = {2049-9647},
  doi = {10.5334/jors.188},
  url = {http://openresearchsoftware.metajnl.com/articles/10.5334/jors.188/},
  urldate = {2020-07-19},
  archiveprefix = {arXiv},
  langid = {english}
}

@article{brownSynergiesSectorCoupling2018,
  title = {Synergies of Sector Coupling and Transmission Reinforcement in a Cost-Optimised, Highly Renewable {{European}} Energy System},
  author = {Brown, T. and Schlachtberger, D. and Kies, A. and Schramm, S. and Greiner, M.},
  year = {2018},
  month = oct,
  journal = {Energy},
  volume = {160},
  pages = {720--739},
  issn = {0360-5442},
  doi = {10.1016/j.energy.2018.06.222},
  url = {https://www.sciencedirect.com/science/article/pii/S036054421831288X},
  urldate = {2023-03-13},
  langid = {english}
}

@article{burandtDecarbonizingChinaEnergy2019,
  title = {Decarbonizing {{China}}'s Energy System -- {{Modeling}} the Transformation of the Electricity, Transportation, Heat, and Industrial Sectors},
  author = {Burandt, Thorsten and Xiong, Bobby and L{\"o}ffler, Konstantin and Oei, Pao-Yu},
  year = {2019},
  month = dec,
  journal = {Applied Energy},
  volume = {255},
  pages = {113820},
  issn = {0306-2619},
  doi = {10.1016/j.apenergy.2019.113820},
  url = {https://www.sciencedirect.com/science/article/pii/S0306261919315077},
  urldate = {2023-02-01},
  langid = {english}
}

@article{caglayanTechnicalPotentialSalt2020,
  title = {Technical Potential of Salt Caverns for Hydrogen Storage in {{Europe}}},
  author = {Caglayan, Dilara Gulcin and Weber, Nikolaus and Heinrichs, Heidi U. and Lin{\ss}en, Jochen and Robinius, Martin and Kukla, Peter A. and Stolten, Detlef},
  year = {2020},
  month = feb,
  journal = {International Journal of Hydrogen Energy},
  volume = {45},
  number = {11},
  pages = {6793--6805},
  issn = {0360-3199},
  doi = {10.1016/j.ijhydene.2019.12.161},
  url = {https://www.sciencedirect.com/science/article/pii/S0360319919347299},
  urldate = {2023-10-31}
}

@misc{comcms,
  title = {Towards an Ambitious Industrial Carbon Management for the {{EU}}},
  author = {{European Commission}},
  year = {2024},
  url = {https://eur-lex.europa.eu/legal-content/EN/TXT/?uri=COM:2024:62:FIN},
  howpublished = {European Commission - Communication}
}

@article{damoreOptimalDesignEuropean2021,
  title = {Optimal Design of {{European}} Supply Chains for Carbon Capture and Storage from Industrial Emission Sources Including Pipe and Ship Transport},
  author = {{d'Amore}, Federico and Romano, Matteo Carmelo and Bezzo, Fabrizio},
  year = {2021},
  month = jan,
  journal = {International Journal of Greenhouse Gas Control},
  volume = {109},
  pages = {103372},
  issn = {1750-5836},
  doi = {10.1016/j.ijggc.2021.103372},
  url = {https://www.sciencedirect.com/science/article/pii/S1750583621001249}
}

@techreport{danishenergyagencyTechnologyDataGeneration2019,
  title = {Technology {{Data}} for {{Generation}} of {{Electricity}} and {{District Heating}}},
  author = {{Danish Energy Agency} and {Energinet.dk}},
  year = {2019},
  institution = {Danish Energy Agency},
  url = {https://ens.dk/en/our-services/projections-and-models/technology-data/technology-data-generation-electricity-and},
  urldate = {2022-08-21},
  langid = {english}
}

@misc{DirectiveEU20232023,
  title = {Directive ({{EU}}) 2023/2413 of the {{European Parliament}} and of the {{Council}} of 18~{{October}} 2023 Amending {{Directive}} ({{EU}})~2018/2001, {{Regulation}} ({{EU}})~2018/1999 and {{Directive}}~98/70/{{EC}} as Regards the Promotion of Energy from Renewable Sources, and Repealing {{Council Directive}} ({{EU}})~2015/652},
  year = {2023},
  month = oct,
  url = {http://data.europa.eu/eli/dir/2023/2413/oj/eng},
  urldate = {2023-12-21},
  langid = {english}
}

@article{eeaCorineLandCover2012,
  title = {Corine {{Land Cover}} ({{CLC}}) 2012, Version 18.5.1},
  author = {{EEA}},
  year = {2012},
  publisher = {European Environment Agency},
  url = {https://land.copernicus.eu/pan-european/corine-land-cover/clc-2012}
}

@article{eeaNatura2000Data2016,
  title = {Natura 2000 Data - the {{European}} Network of Protected Sites},
  author = {{EEA}},
  year = {2016},
  publisher = {European Environment Agency},
  url = {http://www.eea.europa.eu/data-and-maps/data/natura-7}
}

@article{elahiMultiperiodLeastCost2014,
  title = {Multi-Period {{Least Cost Optimisation Model}} of an {{Integrated Carbon Dioxide Capture Transportation}} and {{Storage Infrastructure}} in the {{UK}}},
  author = {Elahi, Nasim and Shah, Nilay and Korre, Anna and Durucan, Sevket},
  year = {2014},
  month = jan,
  journal = {Energy Procedia},
  series = {12th {{International Conference}} on {{Greenhouse Gas Control Technologies}}, {{GHGT-12}}},
  volume = {63},
  pages = {2655--2662},
  issn = {1876-6102},
  doi = {10.1016/j.egypro.2014.11.288},
  url = {https://www.sciencedirect.com/science/article/pii/S1876610214021031},
  urldate = {2023-03-13},
  langid = {english}
}

@techreport{enspreso_database,
  title = {{{ENSPRESO}} - an Open Data, {{EU-28}} Wide, Transparent and Coherent Database of Wind, Solar and Biomass Energy Potentials},
  author = {European Commission, Joint Research Centre (JRC)},
  year = {2019},
  institution = {European Commission, Joint Research Centre (JRC)},
  url = {http://data.europa.eu/89h/74ed5a04-7d74-4807-9eab-b94774309d9f}
}

@misc{eu2023netzero,
  title = {Net-Zero Industry Act},
  author = {{European Commission}},
  year = {2023},
  month = mar,
  url = {https://single-market-economy.ec.europa.eu/industry/sustainability/net-zero-industry-act\_en},
  howpublished = {European Commission - Press release}
}

@misc{europeancommissionEuropeanCO2Storage,
  title = {European {{CO2}} Storage Database},
  author = {{European Commission}},
  url = {https://setis.ec.europa.eu/european-co2-storage-database\_en},
  urldate = {2023-12-12},
  langid = {english}
}

@misc{europeangreendeal,
  title = {The {{European Green Deal}}},
  author = {{European Commission}},
  year = {2019},
  url = {https://ec.europa.eu/info/strategy/priorities-2019-2024/european-green-deal\_en},
  howpublished = {European Commission - Communication}
}

@misc{europeaninnovationfund,
  title = {European Innovation Fund},
  author = {{European Commission}},
  year = {2023},
  url = {https://ec.europa.eu/clima/policies/innovation-fund\_en},
  howpublished = {European Commission - Funding Program}
}

@misc{EurostatCompleteEnergyBalance,
  title = {Complete Energy Balance},
  author = {{Eurostat}},
  year = {2023},
  url = {https://ec.europa.eu/eurostat/databrowser/view/nrg\_bal\_c/default/table?lang=en}
}

@article{friedmann2020net,
  title = {Net-Zero and Geospheric Return: Actions Today for 2030 and Beyond},
  author = {Friedmann, Julio and Zapantis, Alex and Page, Brad and Consoli, Chris and Fan, Zhiyuan and Havercroft, Ian and Liu, Harry and Ochu, Emeka and Raji, Nabeela and Rassool, Dominic and others},
  year = {2020},
  journal = {Center on Global Energy Policy}
}

@misc{gasforclimateEuropeanHydrogenBackbone2022,
  title = {European {{Hydrogen Backbone}} - {{A European Hydrogen Infrastructure Vision Coverving}} 28 {{Countries}}},
  author = {{Gas for Climate}},
  year = {2022},
  month = apr,
  url = {https://gasforclimate2050.eu/wp-content/uploads/2022/04/EHB-A-European-hydrogen-infrastructure-vision-covering-28-countries.pdf}
}

@misc{H2InfrastructureMap,
  title = {H2 {{Infrastructure Map Europe}}},
  url = {https://www.h2inframap.eu/},
  urldate = {2023-11-17},
  langid = {american}
}

@article{hersbachERA5GlobalReanalysis2020,
  title = {The {{ERA5}} Global Reanalysis},
  author = {Hersbach, Hans and Bell, Bill and Berrisford, Paul and Hirahara, Shoji and Hor{\'a}nyi, Andr{\'a}s and {Mu{\~n}oz-Sabater}, Joaqu{\'i}n and Nicolas, Julien and Peubey, Carole and Radu, Raluca and Schepers, Dinand and Simmons, Adrian and Soci, Cornel and Abdalla, Saleh and Abellan, Xavier and Balsamo, Gianpaolo and Bechtold, Peter and Biavati, Gionata and Bidlot, Jean and Bonavita, Massimo and De Chiara, Giovanna and Dahlgren, Per and Dee, Dick and Diamantakis, Michail and Dragani, Rossana and Flemming, Johannes and Forbes, Richard and Fuentes, Manuel and Geer, Alan and Haimberger, Leo and Healy, Sean and Hogan, Robin J. and H{\'o}lm, El{\'i}as and Janiskov{\'a}, Marta and Keeley, Sarah and Laloyaux, Patrick and Lopez, Philippe and Lupu, Cristina and Radnoti, Gabor and {de Rosnay}, Patricia and Rozum, Iryna and Vamborg, Freja and Villaume, Sebastien and Th{\'e}paut, Jean-No{\"e}l},
  year = {2020},
  journal = {Quarterly Journal of the Royal Meteorological Society},
  volume = {146},
  number = {730},
  pages = {1999--2049},
  issn = {1477-870X},
  doi = {10.1002/qj.3803},
  url = {https://onlinelibrary.wiley.com/doi/abs/10.1002/qj.3803},
  urldate = {2023-10-30},
  copyright = {{\copyright} 2020 The Authors. Quarterly Journal of the Royal Meteorological Society published by John Wiley \& Sons Ltd on behalf of the Royal Meteorological Society.},
  langid = {english}
}

@article{hofmannAtliteLightweightPython2021,
  title = {Atlite: {{A Lightweight Python Package}} for {{Calculating Renewable Power Potentials}} and {{Time Series}}},
  shorttitle = {Atlite},
  author = {Hofmann, Fabian and Hampp, Johannes and Neumann, Fabian and Brown, Tom and H{\"o}rsch, Jonas},
  year = {2021},
  month = jun,
  journal = {Journal of Open Source Software},
  volume = {6},
  number = {62},
  pages = {3294},
  issn = {2475-9066},
  doi = {10.21105/joss.03294},
  url = {https://joss.theoj.org/papers/10.21105/joss.03294},
  urldate = {2021-10-20}
}

@inproceedings{hofmannDesigningCO2Network2023,
  title = {Designing a {{CO2 Network}} for a {{Carbon-Neutral European Economy}}},
  booktitle = {2023 19th {{International Conference}} on the {{European Energy Market}} ({{EEM}})},
  author = {Hofmann, Fabian and Tries, Christoph and Neumann, Fabian and Zeyen, Lisa and Brown, Tom},
  year = {2023},
  month = jun,
  pages = {1--7},
  issn = {2165-4093},
  doi = {10.1109/EEM58374.2023.10161887}
}

@article{hofmannLinopyLinearOptimization2023,
  title = {Linopy: {{Linear}} Optimization with n-Dimensional Labeled Variables},
  shorttitle = {Linopy},
  author = {Hofmann, Fabian},
  year = {2023},
  month = apr,
  journal = {Journal of Open Source Software},
  volume = {8},
  number = {84},
  pages = {4823},
  issn = {2475-9066},
  doi = {10.21105/joss.04823},
  url = {https://joss.theoj.org/papers/10.21105/joss.04823},
  urldate = {2023-05-15},
  copyright = {All rights reserved}
}

@article{horschPyPSAEurOpenOptimisation2018,
  title = {{{PyPSA-Eur}}: {{An}} Open Optimisation Model of the {{European}} Transmission System},
  shorttitle = {{{PyPSA-Eur}}},
  author = {H{\"o}rsch, Jonas and Hofmann, Fabian and Schlachtberger, David and Brown, Tom},
  year = {2018},
  month = nov,
  journal = {Energy Strategy Reviews},
  volume = {22},
  pages = {207--215},
  issn = {2211467X},
  doi = {10.1016/j.esr.2018.08.012},
  url = {https://linkinghub.elsevier.com/retrieve/pii/S2211467X18300804},
  urldate = {2019-04-01},
  copyright = {All rights reserved},
  langid = {english}
}

@techreport{hotmaps_industrial_db,
  title = {D2.3 {{WP2}} Report -- Open Data Set for the {{EU28}}},
  author = {Pezzutto, Simon and Zambotti, Stefano and Croce, Silvia and Zambelli, Pietro and Garegnani, Giulia and Scaramuzzino, Chiara and Pascuas, Ram{\'o}n Pascual and Zubaryeva, Alyona and Haas, Franziska and Exner, Dagmar and Mueller, Andreas and Hartner, Michael and Fleiter, Tobias and Klingler, Anna-Lena and Kuehnbach, Matthias and Manz, Pia and Marwitz, Simon and Rehfeldt, Matthias and Steinbach, Jan and Popovski, Eftim},
  year = {2018},
  institution = {Hotmaps Project},
  url = {https://gitlab.com/hotmaps/industrial\_sites/industrial\_sites\_Industrial\_Database}
}

@book{instituteforenergyandtransportjointresearchcentreJRCEUTIMESModelBioenergy2015,
  title = {The {{JRC-EU-TIMES}} Model: Bioenergy Potentials for {{EU}} and Neighbouring Countries},
  shorttitle = {The {{JRC-EU-TIMES}} Model},
  author = {{Institute for Energy {and} Transport (Joint Research Centre)} and Elbersen, Berien and Dalla Longa, Francesco and Ruiz, Pablo and Thiel, Christian and Sgobbi, Alessandra and Hengeveld, Geerten and Kober, Tom and Nijs, Wouter Nijs},
  year = {2015},
  publisher = {Publications Office of the European Union},
  address = {LU},
  url = {https://data.europa.eu/doi/10.2790/39014},
  urldate = {2023-10-31},
  isbn = {978-92-79-53879-7},
  langid = {english},
  lccn = {LD-NA-27-575-EN-N}
}

@techreport{jrc2024,
  title = {Shaping the Future {{CO}}{$_2$} Transport Network for {{Europe}}},
  author = {Tumara, D. and Uihlein, A. and Hidalgo Gonzalez, I.},
  year = {2024},
  pages = {89},
  institution = {European Commission, Joint Research Center},
  url = {https://data.europa.eu/doi/10.2760/582433}
}

@article{lauerCrucialRoleBioenergy2023,
  title = {The {{Crucial Role}} of {{Bioenergy}} in a {{Climate}}-{{Neutral Energy System}} in {{Germany}}},
  author = {Lauer, Markus and Dotzauer, Martin and Millinger, Markus and Oehmichen, Katja and Jordan, Matthias and Kalcher, Jasmin and Majer, Stefan and Thraen, Daniela},
  year = {2023},
  month = mar,
  journal = {Chemical Engineering \& Technology},
  volume = {46},
  number = {3},
  pages = {501--510},
  issn = {0930-7516, 1521-4125},
  doi = {10.1002/ceat.202100263},
  url = {https://onlinelibrary.wiley.com/doi/10.1002/ceat.202100263},
  urldate = {2023-12-15},
  langid = {english}
}

@article{liuExperimentalStudyLeakage2023,
  title = {Experimental Study on the Leakage Temperature Field of Buried {{CO2}} Pipelines},
  author = {Liu, Zhenyi and Xiu, Zihao and Zhao, Yao and Li, Mingzhi and Li, Pengliang and Cai, Peng and Liang, Yizhen},
  year = {2023},
  month = jun,
  journal = {Environmental Science and Pollution Research},
  volume = {30},
  number = {27},
  pages = {70288--70302},
  issn = {1614-7499},
  doi = {10.1007/s11356-023-27289-3},
  url = {https://doi.org/10.1007/s11356-023-27289-3},
  urldate = {2024-07-14},
  langid = {english}
}

@article{lockwoodEuropeanStrategyCarbon,
  title = {A {{European Strategy}} for {{Carbon Capture}} and {{Storage}}},
  author = {Lockwood, Toby and Bertels, Tim},
  langid = {english}
}

@article{mantzosJRCIDEES20152018,
  title = {{{JRC-IDEES}} 2015},
  author = {Mantzos, Leonidas and Matei, Nicoleta Anca and Mulholland, Eamonn and R{\'o}zsai, M{\'a}t{\'e} and Tamba, Marie and Wiesenthal, Tobias},
  year = {2018},
  month = jun,
  publisher = {European Commission, Joint Research Centre (JRC)},
  doi = {10.2905/JRC-10110-10001},
  url = {http://data.europa.eu/89h/jrc-10110-10001},
  urldate = {2023-11-14},
  langid = {english}
}

@techreport{mcvayreneeMethaneEmissionsGas2023,
  title = {Methane {{Emissions From U}}.{{S}}. {{Gas Pipeline Leaks}}},
  author = {{McVay, Renee}},
  year = {2023},
  month = aug,
  institution = {Environmental Defense Fund},
  url = {https://www.edf.org/sites/default/files/documents/Pipeline\%20Methane\%20Leaks\%20Report.pdf},
  urldate = {2024-08-11}
}

@article{middletonSimCCSOpensourceTool2020,
  title = {{{SimCCS}}: {{An}} Open-Source Tool for Optimizing {{CO2}} Capture, Transport, and Storage Infrastructure},
  shorttitle = {{{SimCCS}}},
  author = {Middleton, Richard S. and Yaw, Sean P. and Hoover, Brendan A. and Ellett, Kevin M.},
  year = {2020},
  month = feb,
  journal = {Environmental Modelling \& Software},
  volume = {124},
  pages = {104560},
  issn = {1364-8152},
  doi = {10.1016/j.envsoft.2019.104560},
  url = {https://www.sciencedirect.com/science/article/pii/S1364815218300185},
  urldate = {2022-11-29},
  langid = {english}
}

@misc{millingerDiversityBiomassUsage2023,
  title = {Diversity of Biomass Usage Pathways to Achieve Emissions Targets in the {{European}} Energy System},
  author = {Millinger, Markus and Hedenus, Fredrik and Reichenberg, Lina and Zeyen, Elisabeth and Neumann, Fabian and Berndes, G{\"o}ran},
  year = {2023},
  month = jul,
  issn = {2693-5015},
  doi = {10.21203/rs.3.rs-3097648/v1},
  url = {https://www.researchsquare.com/article/rs-3097648/v1},
  urldate = {2024-09-09}
}

@article{morbeeOptimisedDeploymentEuropean2012,
  title = {Optimised Deployment of a {{European CO2}} Transport Network},
  author = {Morbee, Joris and Serpa, Joana and Tzimas, Evangelos},
  year = {2012},
  month = mar,
  journal = {International Journal of Greenhouse Gas Control},
  volume = {7},
  pages = {48--61},
  issn = {1750-5836},
  doi = {10.1016/j.ijggc.2011.11.011},
  url = {https://www.sciencedirect.com/science/article/pii/S1750583611002210},
  urldate = {2023-03-13},
  langid = {english}
}

@misc{muehlenpfordtTimeSeries2019,
  title = {Time Series},
  author = {Muehlenpfordt, Jonathan},
  year = {2019},
  month = jun,
  publisher = {Open Power System Data},
  doi = {10.25832/TIME_SERIES/2019-06-05},
  url = {https://data.open-power-system-data.org/time\_series/2019-06-05},
  urldate = {2023-11-14}
}

@data{NationalEmissionsReported2023,
  title = {National Emissions Reported to the {{UNFCCC}} and to the {{EU}} Greenhouse Gas Monitoring Mechanism, October 2023},
  year = {2023},
  publisher = {European Environment Agency},
  url = {https://sdi.eea.europa.eu/catalogue/srv/api/records/e2e7dd1e-0d67-4b20-a0d4-b22c53a59d24}
}

@misc{NaturalGasTransmission2021,
  title = {Natural Gas Transmission Leakage Rates},
  year = {2021},
  month = apr,
  journal = {Global Energy Monitor},
  url = {https://www.gem.wiki/Natural\_gas\_transmission\_leakage\_rates},
  urldate = {2024-08-11},
  langid = {english}
}

@misc{neumannBenefitsHydrogenNetwork2022,
  title = {Benefits of a {{Hydrogen Network}} in {{Europe}}},
  author = {Neumann, Fabian and Zeyen, Elisabeth and Victoria, Marta and Brown, Tom},
  year = {2022},
  month = jul,
  number = {arXiv:2207.05816},
  eprint = {2207.05816},
  primaryclass = {physics},
  publisher = {arXiv},
  url = {http://arxiv.org/abs/2207.05816},
  urldate = {2022-11-29},
  archiveprefix = {arXiv}
}

@misc{neumannEnergyImportsInfrastructure2024,
  title = {Energy {{Imports}} and {{Infrastructure}} in a {{Carbon-Neutral European Energy System}}},
  author = {Neumann, Fabian and Hampp, Johannes and Brown, Tom},
  year = {2024},
  month = apr,
  journal = {arXiv.org},
  url = {https://arxiv.org/abs/2404.03927v1},
  urldate = {2024-09-06},
  langid = {english}
}

@misc{NorthernLightsWhat,
  title = {Northern {{Lights}} -- {{What}} We Do},
  url = {https://norlights.com/what-we-do/},
  urldate = {2023-11-09},
  langid = {american}
}

@article{oeiModelingCarbonCapture2014,
  title = {Modeling a {{Carbon Capture}}, {{Transport}}, and {{Storage Infrastructure}} for {{Europe}}},
  author = {Oei, Pao-Yu and Herold, Johannes and Mendelevitch, Roman},
  year = {2014},
  month = dec,
  journal = {Environmental Modeling \& Assessment},
  volume = {19},
  number = {6},
  pages = {515--531},
  issn = {1573-2967},
  doi = {10.1007/s10666-014-9409-3},
  url = {https://doi.org/10.1007/s10666-014-9409-3},
  urldate = {2023-03-13},
  langid = {english}
}

@misc{pfeifrothSurfaceRadiationData2017,
  title = {Surface {{Radiation Data Set}} - {{Heliosat}} ({{SARAH}}) - {{Edition}} 2},
  author = {Pfeifroth, Uwe and Kothe, Steffen and M{\"u}ller, Richard and Trentmann, J{\"o}rg and Hollmann, Rainer and Fuchs, Petra and Werscheck, Martin},
  year = {2017},
  month = jun,
  pages = {7.1 TiB},
  publisher = {Satellite Application Facility on Climate Monitoring (CM SAF)},
  doi = {10.5676/EUM_SAF_CM/SARAH/V002},
  url = {https://wui.cmsaf.eu/safira/action/viewDoiDetails?acronym=SARAH\_V002},
  urldate = {2023-10-30}
}

@misc{piamanzGeoreferencedIndustrialSites2018,
  title = {Georeferenced Industrial Sites with Fuel Demand and Excess Heat Potential},
  author = {Pia Manz, Tobias Fleiter},
  year = {2018},
  month = mar,
  publisher = {Zenodo},
  doi = {10.5281/zenodo.4687147},
  url = {https://zenodo.org/record/4687147},
  urldate = {2023-03-03}
}

@article{pickeringDiversityOptionsEliminate2022,
  title = {Diversity of Options to Eliminate Fossil Fuels and Reach Carbon Neutrality across the Entire {{European}} Energy System},
  author = {Pickering, Bryn and Lombardi, Francesco and Pfenninger, Stefan},
  year = {2022},
  month = jun,
  journal = {Joule},
  volume = {6},
  number = {6},
  pages = {1253--1276},
  publisher = {Elsevier},
  issn = {2542-4785, 2542-4351},
  doi = {10.1016/j.joule.2022.05.009},
  url = {https://www.cell.com/joule/abstract/S2542-4351(22)00236-7},
  urldate = {2023-12-11},
  langid = {english}
}

@misc{PyPSAEurSecSectorCoupledOpen2023,
  title = {{{PyPSA-Eur-Sec}}: {{A Sector-Coupled Open Optimisation Model}} of the {{European Energy System}}.},
  year = {2023},
  url = {https://github.com/pypsa/pypsa-eur-sec}
}

@article{righettiSitingCarbonDioxide2017,
  title = {Siting {{Carbon Dioxide Pipelines}}},
  author = {Righetti, Tara K.},
  year = {2017},
  journal = {ONE J: Oil and Gas, Natural Resources, and Energy Journal},
  volume = {3},
  pages = {907},
  url = {https://heinonline.org/HOL/Page?handle=hein.journals/onej3\&id=927\&div=\&collection=}
}

@article{stewartFeasibilityEuropeanwideIntegrated2014,
  title = {The Feasibility of a {{European-wide}} Integrated {{CO2}} Transport Network},
  author = {Stewart, R. J. and Scott, V. and Haszeldine, R. S. and Ainger, D. and Argent, S.},
  year = {2014},
  journal = {Greenhouse Gases: Science and Technology},
  volume = {4},
  number = {4},
  pages = {481--494},
  issn = {2152-3878},
  doi = {10.1002/ghg.1410},
  url = {https://onlinelibrary.wiley.com/doi/abs/10.1002/ghg.1410},
  urldate = {2023-11-13},
  copyright = {{\copyright} 2014 Society of Chemical Industry and John Wiley \& Sons, Ltd.},
  langid = {english}
}

@misc{TESHydrogenLife2023,
  title = {{{TES Hydrogen}} for Life --- {{Green Cycle}}},
  year = {2023},
  url = {https://experience.tes-h2.com/},
  urldate = {2023-02-01},
  langid = {english}
}

@techreport{thedanishenergyagencyTechnologyDataCarbon2023,
  title = {Technology {{Data}} for {{Carbon Capture}}, {{Transport}} and {{Storage}}},
  author = {{The Danish Energy Agency}},
  year = {2023},
  month = nov,
  url = {https://ens.dk/sites/ens.dk/files/Analyser/technology\_data\_for\_carbon\_capture\_transport\_and\_storage.pdf},
  urldate = {2023-11-27}
}

@misc{ToolsGreenTransition,
  title = {Tools for the Green Transition},
  journal = {Endrava},
  url = {https://www.endrava.no/?lang=en},
  urldate = {2022-11-29},
  langid = {american}
}

@misc{uwekrienDemandlib2023,
  title = {Demandlib},
  author = {{Uwe Krien} and {Patrick Sch{\"o}nfeldt} and {et al.}},
  year = {2023},
  month = sep,
  url = {https://github.com/oemof/demandlib},
  urldate = {2023-11-14},
  copyright = {MIT},
  howpublished = {oemof community}
}

@article{vitaliRisksSafetyCO22021,
  title = {Risks and {{Safety}} of {{CO2 Transport}} via {{Pipeline}}: {{A Review}} of {{Risk Analysis}} and {{Modeling Approaches}} for {{Accidental Releases}}},
  shorttitle = {Risks and {{Safety}} of {{CO2 Transport}} via {{Pipeline}}},
  author = {Vitali, Matteo and Zuliani, Cristina and Corvaro, Francesco and Marchetti, Barbara and Terenzi, Alessandro and Tallone, Fabrizio},
  year = {2021},
  month = jan,
  journal = {Energies},
  volume = {14},
  number = {15},
  pages = {4601},
  publisher = {Multidisciplinary Digital Publishing Institute},
  issn = {1996-1073},
  doi = {10.3390/en14154601},
  url = {https://www.mdpi.com/1996-1073/14/15/4601},
  urldate = {2024-07-14},
  copyright = {http://creativecommons.org/licenses/by/3.0/},
  langid = {english}
}

@article{weiProposedGlobalLayout2021,
  title = {A Proposed Global Layout of Carbon Capture and Storage in Line with a 2 {$^\circ$}{{C}} Climate Target},
  author = {Wei, Yi-Ming and Kang, Jia-Ning and Liu, Lan-Cui and Li, Qi and Wang, Peng-Tao and Hou, Juan-Juan and Liang, Qiao-Mei and Liao, Hua and Huang, Shi-Feng and Yu, Biying},
  year = {2021},
  month = feb,
  journal = {Nature Climate Change},
  volume = {11},
  number = {2},
  pages = {112--118},
  publisher = {Nature Publishing Group},
  issn = {1758-6798},
  doi = {10.1038/s41558-020-00960-0},
  url = {https://www.nature.com/articles/s41558-020-00960-0},
  urldate = {2022-11-29},
  copyright = {2021 The Author(s), under exclusive licence to Springer Nature Limited},
  langid = {english}
}

@misc{WeTurnCO2,
  title = {We Turn {{CO2}} into Stone - {{Carbfix}}},
  url = {https://www.carbfix.com/},
  urldate = {2023-03-20},
  langid = {english}
}

@misc{wiegmansGridkitExtractEntsoE2016,
  title = {Gridkit {{Extract Of Entso-E Interactive Map}}},
  author = {Wiegmans, Bart},
  year = {2016},
  month = jun,
  publisher = {Zenodo},
  doi = {10.5281/ZENODO.55853},
  url = {https://zenodo.org/record/55853},
  urldate = {2020-07-19},
  copyright = {Creative Commons Attribution 4.0, Open Access}
}

@misc{zeyenPyPSATechnologydataTechnology2023,
  title = {{{PyPSA}}/Technology-Data: {{Technology Data}} v0.5.0},
  shorttitle = {{{PyPSA}}/Technology-Data},
  author = {Zeyen, Lisa and Hammp, Johannes and Millinger, Markus and Neumann, Fabian and Brown, Tom and Martavp and Parzen, Max and Lukasnacken},
  year = {2023},
  month = feb,
  doi = {10.5281/ZENODO.7622914},
  url = {https://zenodo.org/record/7622914},
  urldate = {2023-03-01},
  copyright = {Open Access},
  howpublished = {Zenodo}
}

\newpage
\appendix
\setcounter{section}{0}
\renewcommand{\thesection}{\Alph{section}}
\renewcommand{\thefigure}{\Alph{section}.\arabic{figure}}

\onecolumn % Add this line to adjust the layout to a single column

\section{Appendix}

\subsection{Limitations}
\label{sec:limitations}
While the model is comprehensive, it has certain limitations that affect the validity of its results. Additionally, there are important aspects of the findings that require further context and discussion.
In reality, energy demand and renewable supply can only be estimated for the next few hours/days. The model's perfect foresight may lead to non-reproducible technology dispatch which neglects uncertainties from long-term predictions and necessary reserve margins.
Demand and emissions from industrial clusters heavily drive the modeling results. Allowing the model to relocate these, incorporate flexibility measures or alter underlying processes, may lead to less dependency on \carbon{} and \hydrogen{} networks than with the static industry locations assumed in this study.
The transport of fuel (FT, gas, oil) between the regions is not limited which may overestimate their flexibility. However, the distributed CU systems, as enabled through the hydrogen network, align well with today's distributed locations for oil refineries, providing kerosene for aviation and naphtha for industry. Our assumption to run Fischer-Tropsch facilities on baseload with at least 90\% of their nominal capacity is likely shaping the deployment of CU, underestimating potential important flexibilities.
The assumed cost projections on technologies are subject to uncertainties. As shown in \ref{sec:subsidy}, the effect of a 50\% cost reduction on \carbon{} pipelines alters total system costs to a small extent (below 1\%). However, a different set of costs on DAC and electrolysis may well impact the model results.
Considering different strategies for carbon removal, DAC deployment could be in parts replaced if larger sustainable biomass sources are assumed or imported, as shown in~\cite{lauerCrucialRoleBioenergy2023} and~\cite{millingerDiversityBiomassUsage2023} respectively.

Finally, our energy system model assumes total autarky of the European system and neglects potential imports from elsewhere. Depending on the type of energy carrier and whether it is \carbon{} neutral or not, allowing imports can change the optimal design and operation of the system quite significantly, as discussed in~\cite{neumannEnergyImportsInfrastructure2024}. In our model, we see that when higher sequestration rates are allowed, the model imports more fossil oil, which reduces the use of CU in Central Europe and thus the hydrogen transport from South to Central Europe (see Section~\ref{sec:higher_sequestration}). While this substitution reduces the need for Fischer-Tropsch fuel production, it also requires significant carbon sequestration within Europe to offset the emissions from fossil oil use. If the imported oil were carbon-neutral or green, there would be no need for dedicated sequestration infrastructure as the emissions would already be accounted for. In such a case, the overall energy infrastructure would remain largely the same, minus the additional carbon sequestration capacity needed to offset the emissions from fossil imports. Thus, both the \hydrogen{} and \carbon{} networks would be less relevant.
In the context of green hydrogen imports, as highlighted by~\cite{neumannEnergyImportsInfrastructure2024}, the model is expected to reduce the use of electrolysis and methanolization in southern European countries, as well as hydrogen transport routes from there to central Europe. However, the deployment of CU near point sources of \carbon{} would still be as in our scenario. The same applies to the deployment of a carbon network driven by fixed sequestration sites.
A disadvantage of importing synthetic fuels would be that the significant \carbon{} available in sustainable biomass sources in Europe would be underutilized, and there would be an additional burden of regulating the sourcing of sustainable \carbon{} outside of Europe.

\subsection{Technologies with Carbon Capture}

In the following, we give a more detailed description of technologies that are upgraded to integrate carbon capture.
In the \baselinescenario{}, the capture share is highest for biogas-to-gas facilities, followed by process emissions, biomass for industry, biomass CHPs, gas for industry, SMR and gas CHPs. For models with additional transport systems, we observe a correlation between CC share and capacity factors of the underlying technologies: Process emissions as well as gas and biomass for industry have high capacity factors (above 80\%), and CC shares close to 100\%, the only exception being gas for industry in the \hydrogenscenario{}. Biomass CHPs have a capacity factor of around 50\%, and CC shares of 50\% in the \hydrogenscenario{} and 97\% in the \carbonscenario{}. SMR and gas CHPs have capacity factors around 15\%, and CC shares at nearly 0\% for the \baselinescenario{}. The capacity factor and CC share of SMR increase to around 20\% and 50ß-65\%, respectively, in the other models. Gas CHPs serve as peak load electricity production technology, and with very few operating hours, thus the high investment costs into CC applications are not efficient. Similarly, SMR serves as a ``peak-load'' hydrogen production technology in regions with poor renewable resources for electrolysis.

\clearpage
\subsection{Sensitivities}
\label{sec:sensitivities}

\subsubsection*{Higher sequestration limits}
\label{sec:higher_sequestration}
\begin{figure}[ht!]
    \centering
    \begin{subfigure}{.5\textwidth}
        \includegraphics[width=\linewidth]{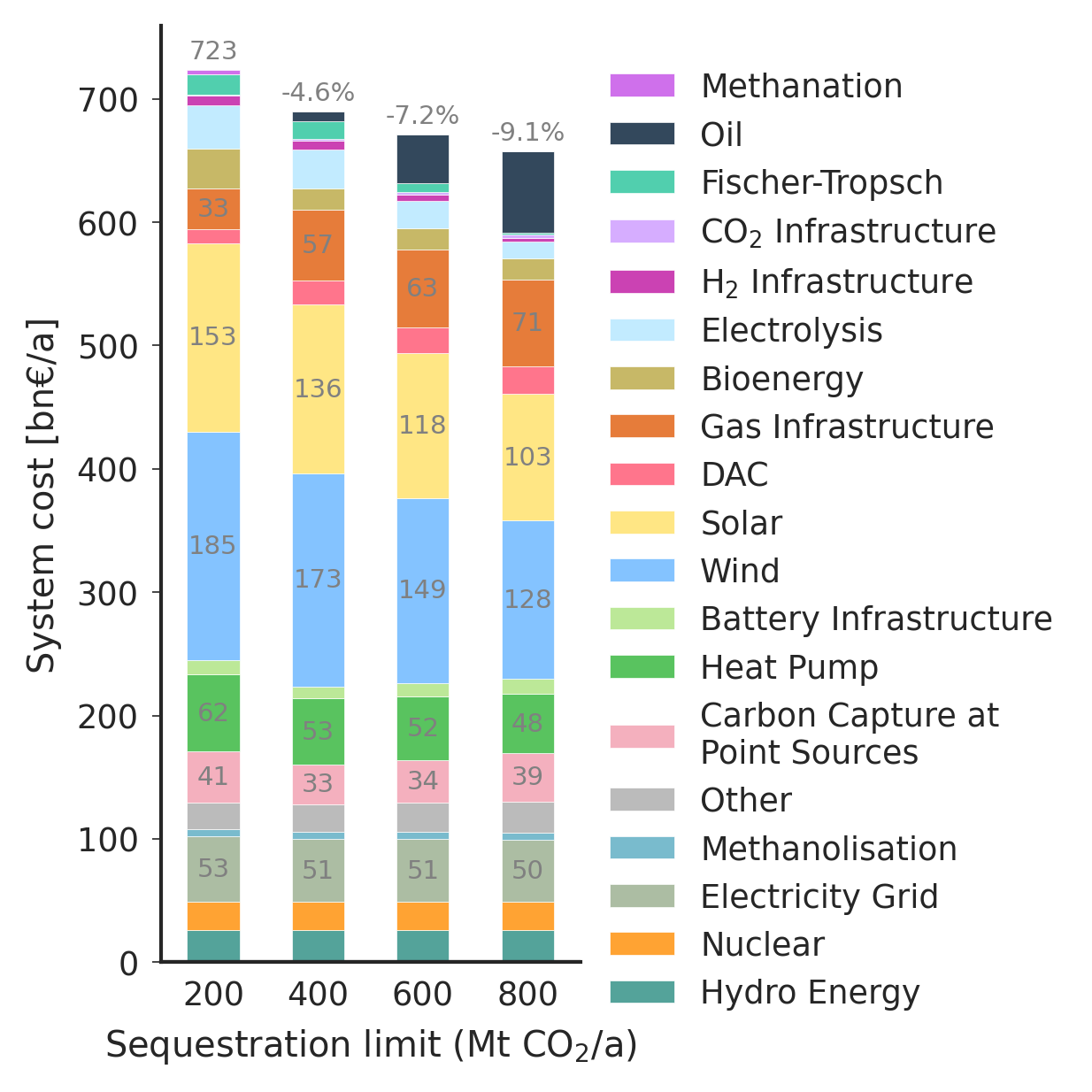}
        \caption[short]{}
        \label{fig:sequestration_cost_bar}
    \end{subfigure}%
    \begin{subfigure}{.5\textwidth}
        \includegraphics[width=\linewidth]{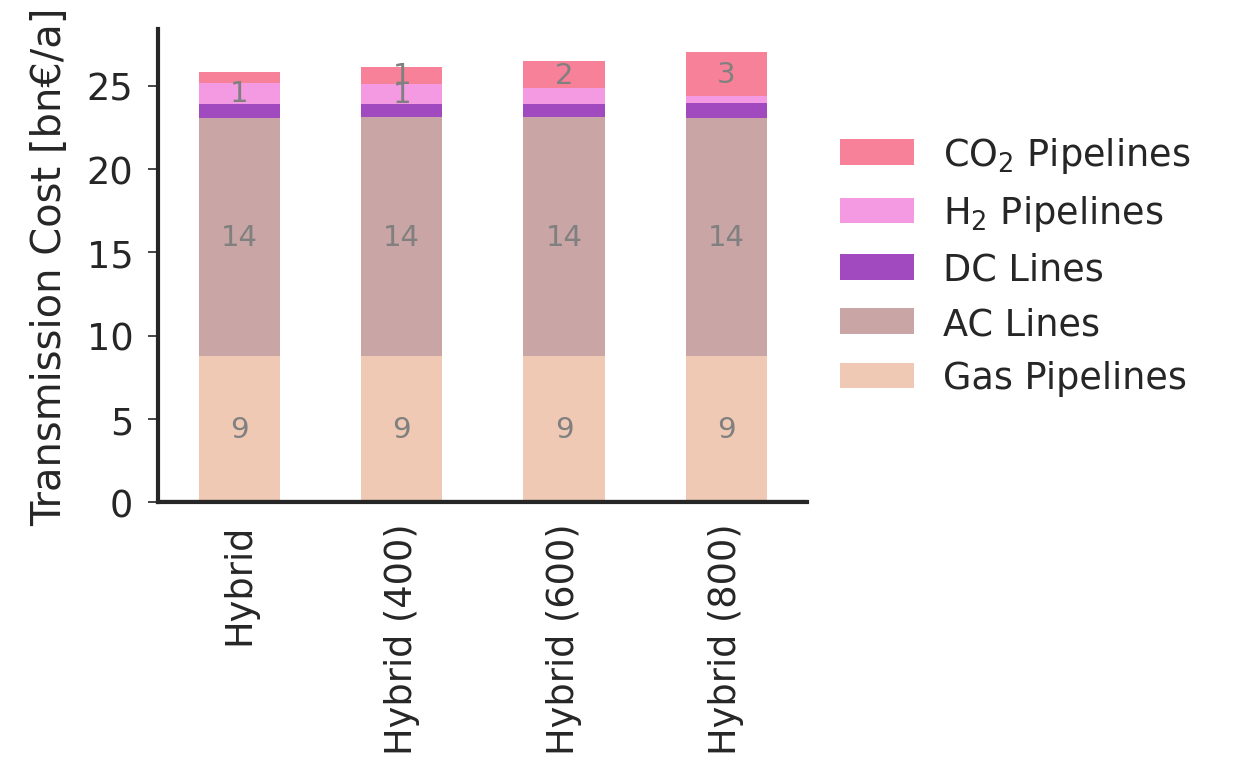}
        \caption[short]{}
        \label{fig:sequestration_cost_bar_transmission}
    \end{subfigure}
    \caption{System costs for the \hybridscenario{} as a function of the sequestration limit (a). The transmission system cost for the \hybridscenario{} as a function of the sequestration limit (b).}
\end{figure}

Carbon management in the optimized energy system, particularly in the hard-to-abate sectors, is heavily influenced by the model's sequestration potential, which is capped at 200~Mt/a. Lifting the sequestration limit leads to cost benefits and a shift in primary energy supply and its implications (see Fig.~\ref{fig:sequestration_cost_bar}). With higher sequestration rates, more fossil fuels are used in the aviation, heating, and industry sectors in land-locked regions where hydrogen is expensive. This reduces the reliance on carbon utilization (CU) and, therefore, the production and transport of hydrogen. At the same time, the carbon network is expanded to transport more carbon from inland point sources to sequestration sites. Sequestration is further supported by large DAC facilities near coastal areas to offset additional emissions from fossil fuels.
These effects lead to significant changes in investments when increasing sequestration from 200~Mt/a to 800~Mt/a. Due to reduced hydrogen production, investments in wind and solar drop by a third. Total system costs are reduced by 9.1\%, primarily due to lower renewable energy investments. Regarding transport networks, hydrogen network capacity is halved, while the carbon network quadruples (see Fig.~\ref{fig:sequestration_cost_bar_transmission}). These findings are consistent with the effects discussed in~\cite{hofmannDesigningCO2Network2023}.

As sequestration capacity increases, the cost savings from additional sequestration begin to plateau. At 200~Mt/a, the shadow price of the sequestration constraint shows that each additional tonne saves 225~€/t, but at 800~Mt/a, this saving drops to 51~€/t. Extrapolating this behavior, saturation of sequestration can be expected around 1200~Mt/a.

While increasing sequestration capacity offers economic benefits, it poses potential risks. Over-reliance on sequestration may delay necessary investments in cleaner technologies and renewable energy. Additionally, the large-scale expansion of carbon storage infrastructure may encounter public opposition due to concerns about the long-term stability of CO$_2$ storage sites and possible leakage risks. Furthermore, the focus on sequestration could perpetuate the use of fossil fuels in sectors where decarbonization alternatives, such as electrification or green hydrogen, may offer more sustainable solutions. In particular, if sequestration sites are overused for compensating continuing unabated fossil emissions, they become unavailable for net-negative targets necessary to compensate historical emissions. Thus, careful consideration must be given to avoid locking in fossil fuel usage and undermining the broader energy transition goals.

\subsubsection{Lower \carbon{} pipelines costs}
\label{sec:subsidy}

\begin{figure}[ht!]
    \centering
    \begin{subfigure}{.5\textwidth}
        \includegraphics[width=\linewidth]{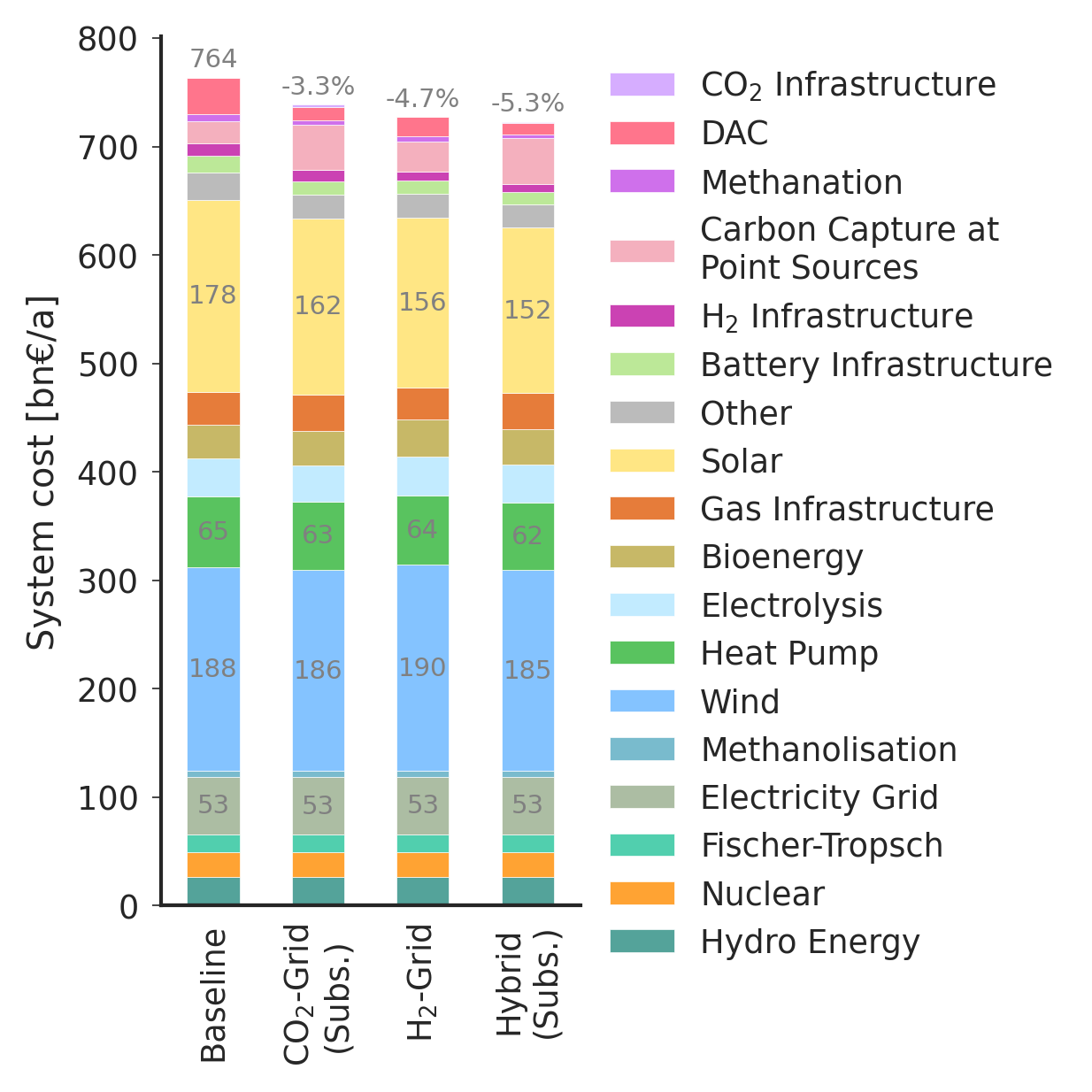}
        \caption{}
        \label{fig:cost_bar_subsidy}
    \end{subfigure}%
    \begin{subfigure}{.5\textwidth}
        \centering
        \includegraphics[width=\linewidth]{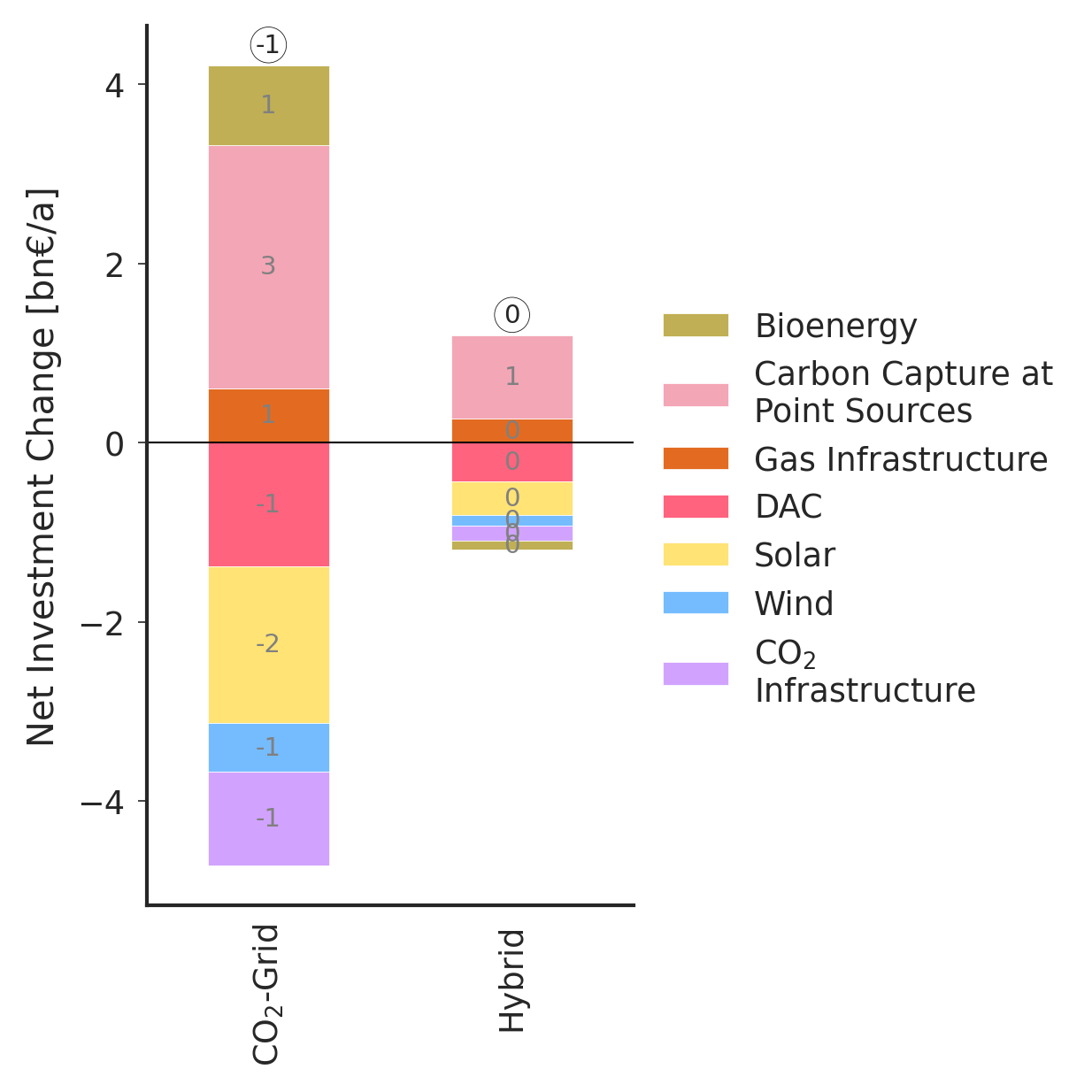}
        \caption{}
        \label{fig:cost_bar_diff_subsidy}
    \end{subfigure}
    \caption{System costs for all models in the net-zero emission scenario with 50\% subsidy of \carbon{} pipelines (a) and change in system cost for the \carbongrid{} and \hybridscenario{} when subsidizing \carbon{} pipelines by 50\% (b).}
\end{figure}

The effect of a 50\% cost-reduction on \carbon{} pipelines, such as through subsidies, impacts the optimal technology deployment to a small extent. Fig.~\ref{fig:cost_bar_subsidy} shows the corresponding total system cost of all scenarios with a net carbon neutrality target. Fig.~\ref{fig:cost_bar_diff_subsidy} shows the net change in system cost between non-subsidized and subsidized models, showing only the \carbongrid{} and \hybridscenario{} where changes occur.

\subsubsection{Higher Electrolysis Costs}

\begin{figure}[ht!]
    \centering
    \includegraphics[width=0.5\linewidth]{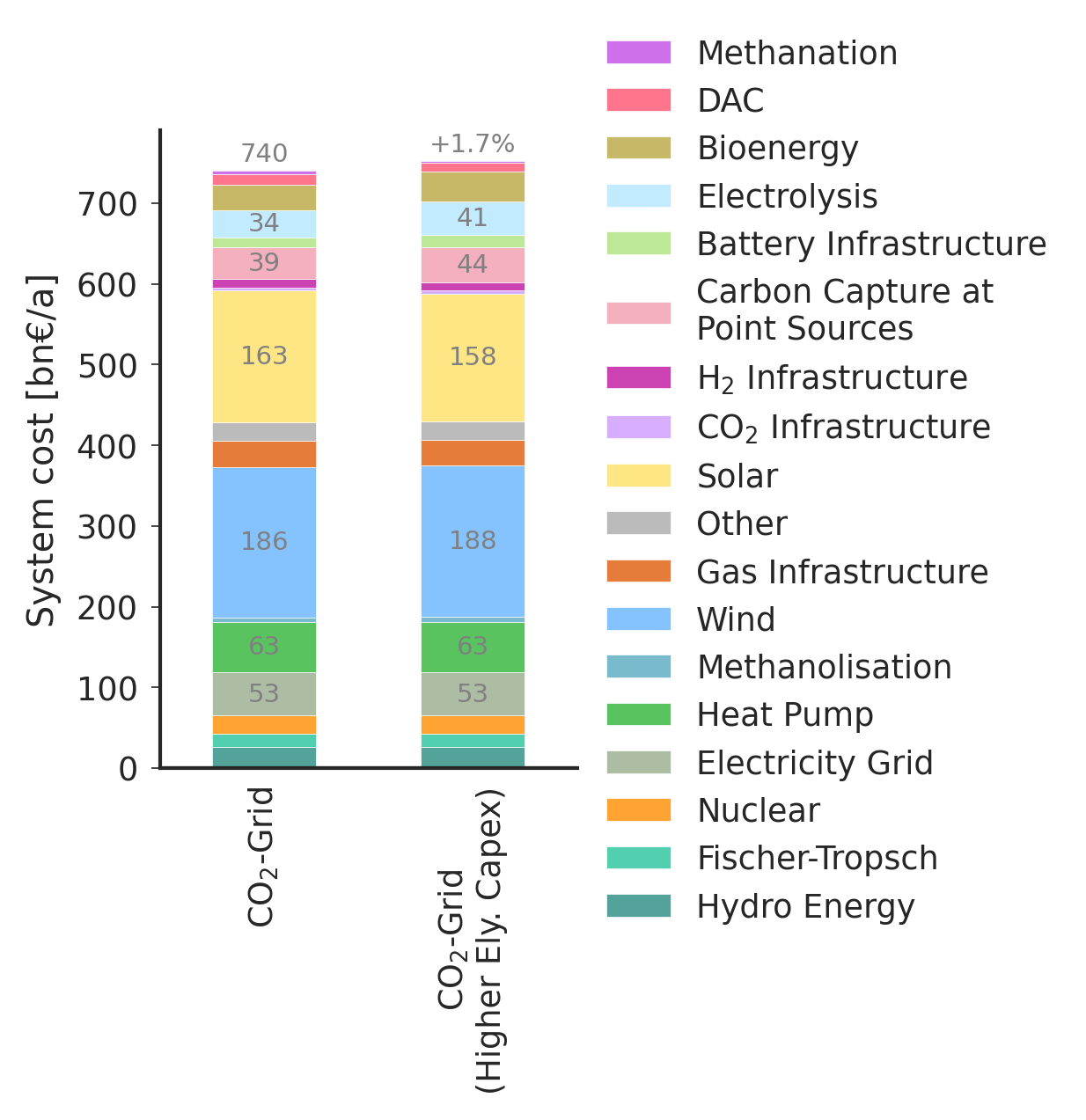}
    \caption{System costs for the \carbonscenario{} when increasing electrolysis investment costs by 50\%.}
    \label{fig:cost_bar_electrolysis_co2}
\end{figure}

\begin{figure}[ht!]
    \centering
    \includegraphics[width=0.5\linewidth]{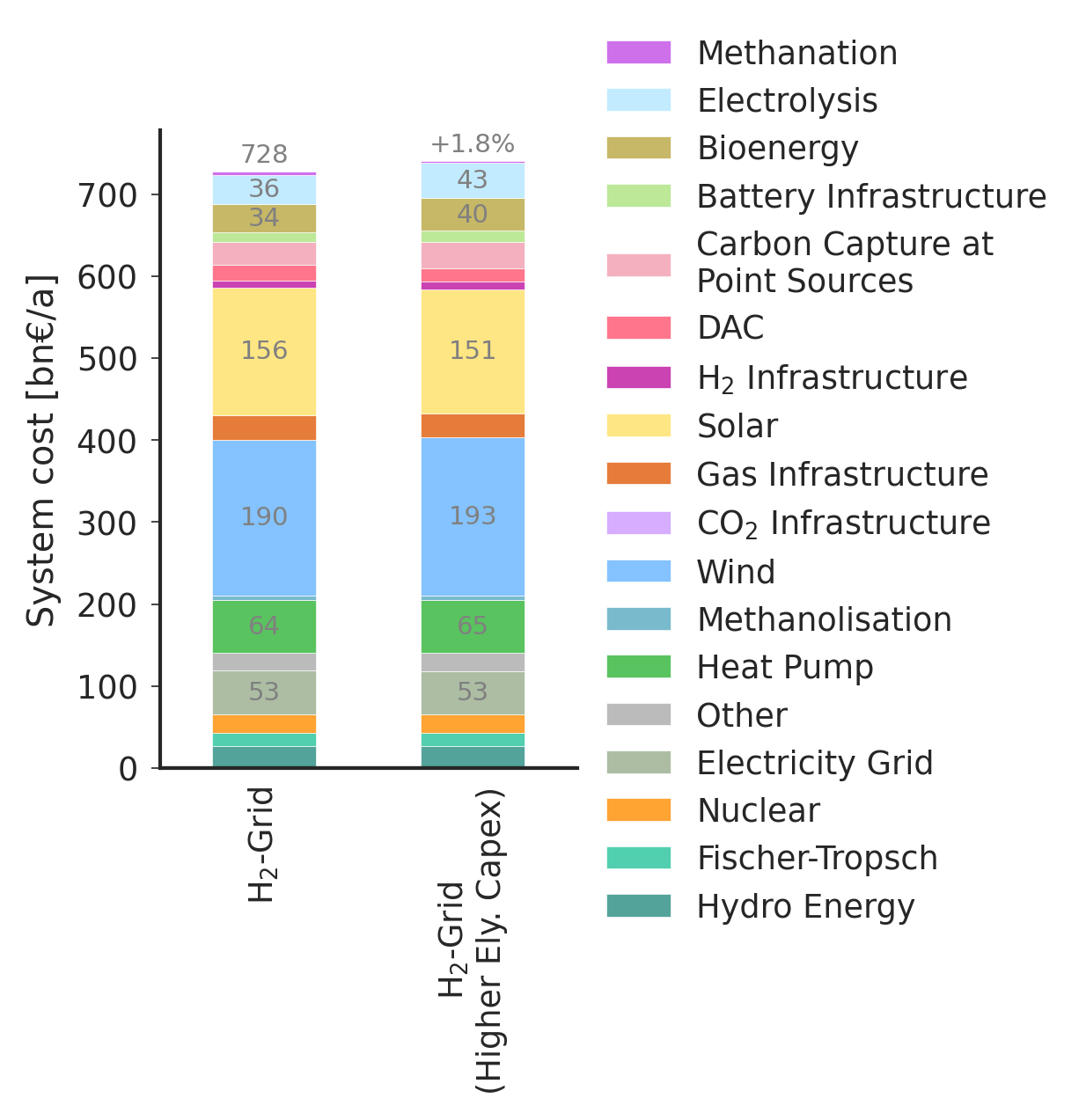}
    \caption{System costs for the \hydrogenscenario{} when increasing electrolysis investment costs by 50\%.}
    \label{fig:cost_bar_electrolysis_h2}
\end{figure}

\begin{figure}[ht!]
    \centering
    \includegraphics[width=0.5\linewidth]{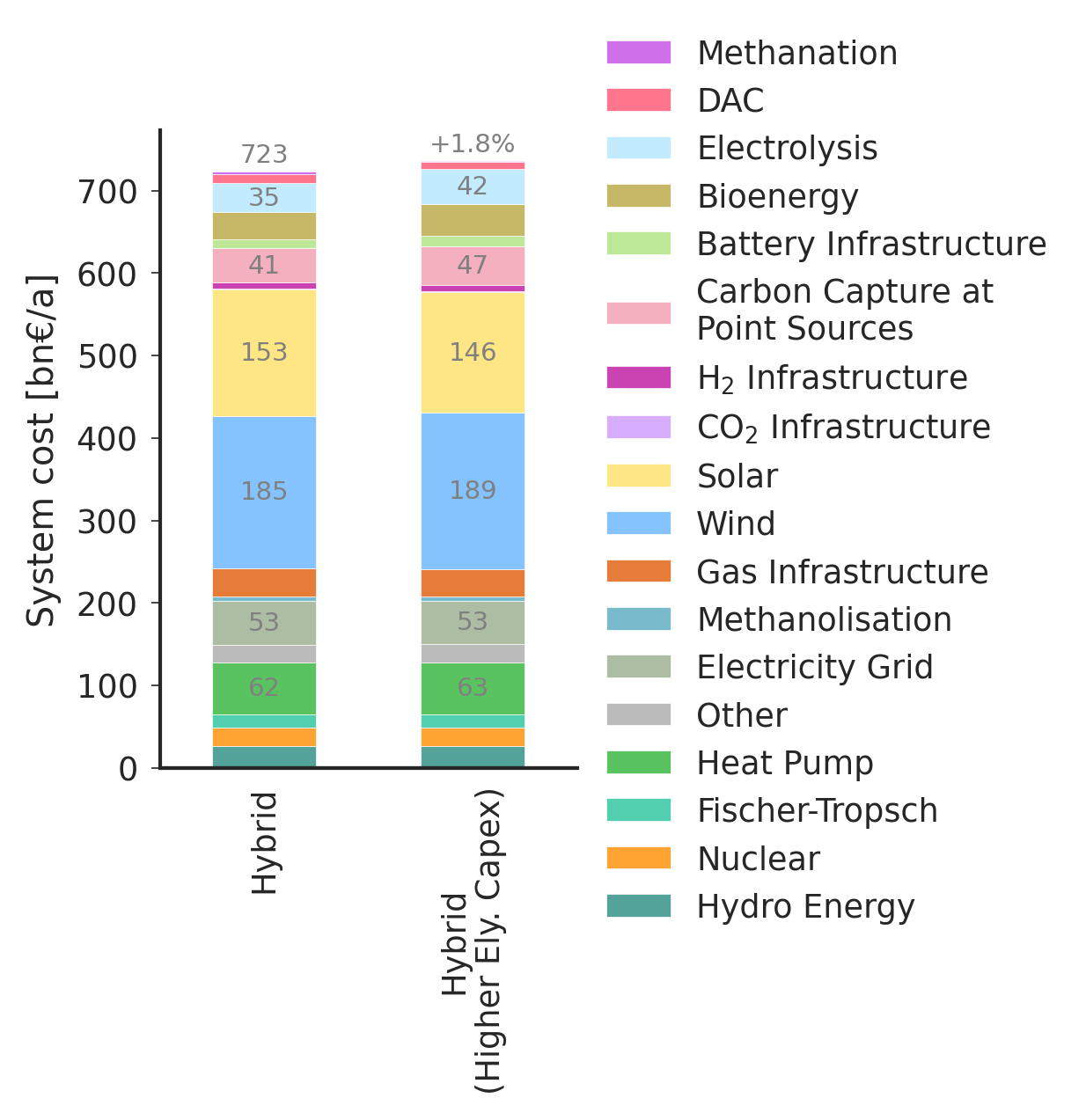}
    \caption{System costs for the \hybridscenario{} when increasing electrolysis investment costs by 50\%.}
    \label{fig:cost_bar_electrolysis_full}
\end{figure}

The effect of a 50\% increase in electrolysis investment costs on the optimal technology deployment is shown in Fig.~\ref{fig:cost_bar_electrolysis_co2} for the \carbonscenario{}, Fig.~\ref{fig:cost_bar_electrolysis_h2} for the \hydrogenscenario{}, and Fig.~\ref{fig:cost_bar_electrolysis_full} for the \hybridscenario{}. The total system costs increase by 1.7\% in the \carbonscenario{}, by 1.8\% in the \hydrogenscenario{} and the \hybridscenario{}.

\clearpage
\subsection{Other sectors in the energy system}

The following maps show the optimal supply, demand, transport and prices of energy carriers in the \hybridscenario{} not covered in the main text.

\begin{figure}[ht!]
    \centering
    \begin{subfigure}{0.5\textwidth}
        \includegraphics[width=\linewidth]{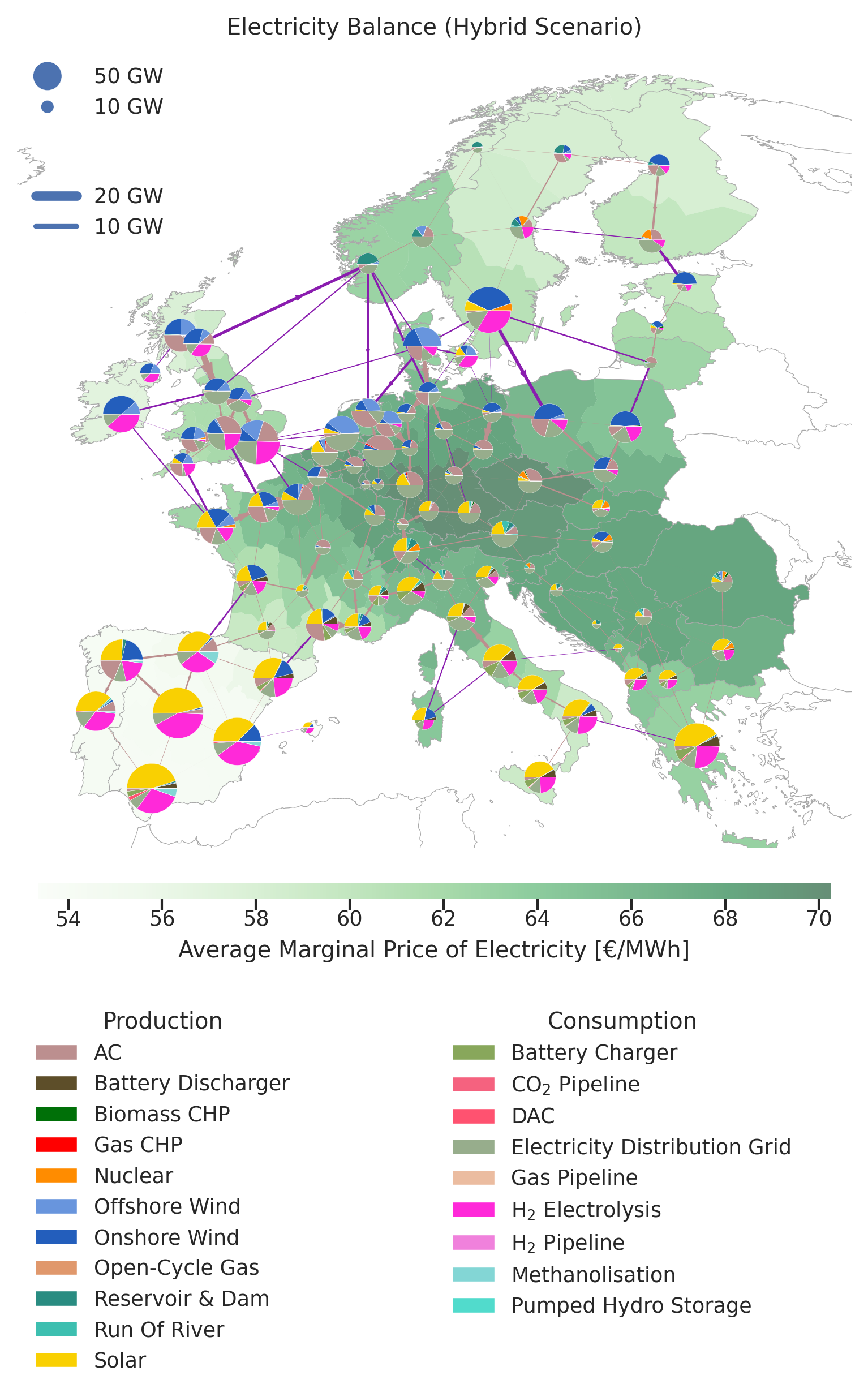}
        \caption{Optimal supply, consumption, transport and prices of electricity in the \hybridscenario{}.}
        \label{fig:balance_map_electricity}
    \end{subfigure}%
    \begin{subfigure}{0.5\textwidth}
        \includegraphics[width=\linewidth]{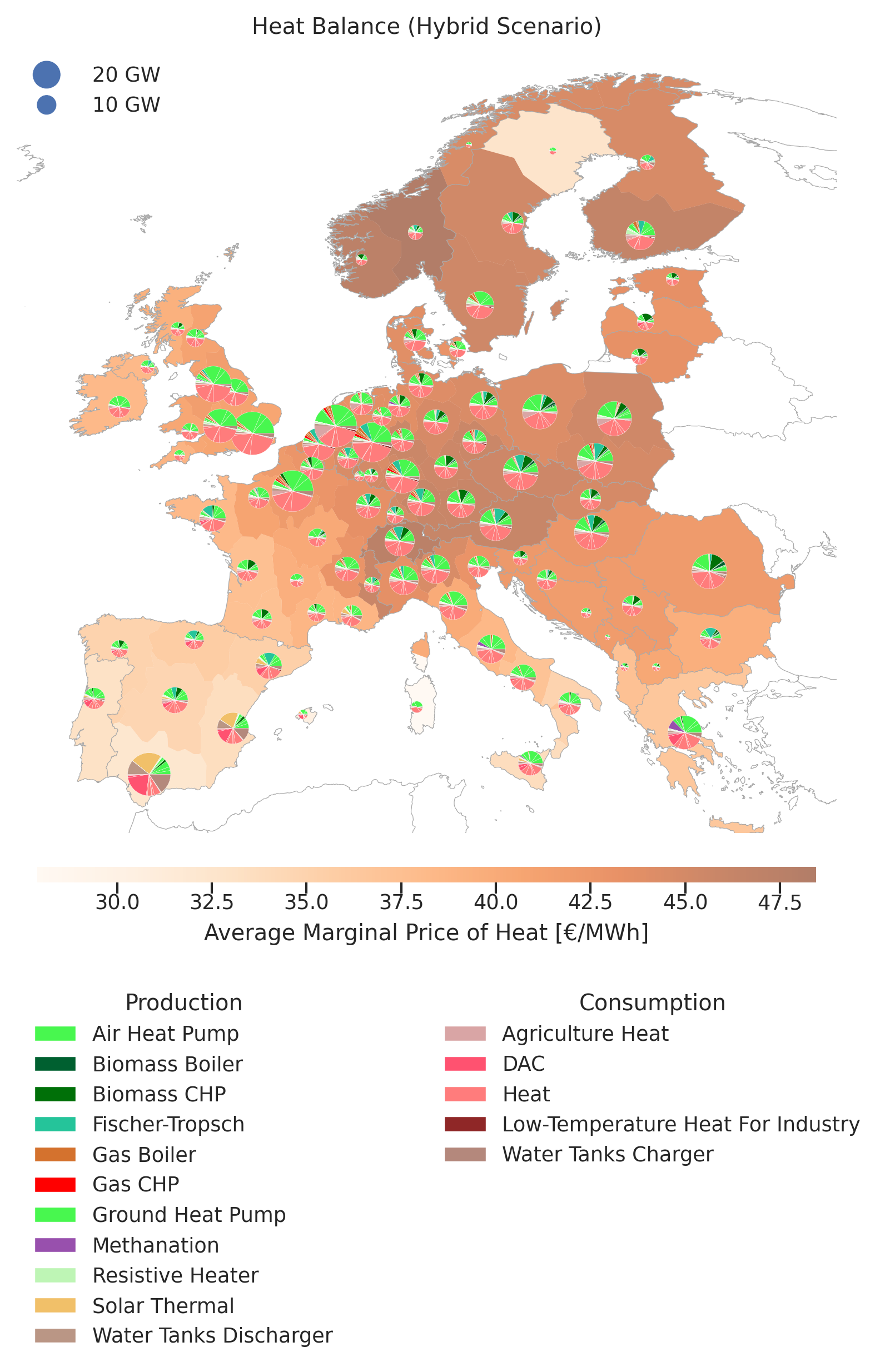}
        \caption{Optimal supply, consumption, transport and prices of heat in the \hybridscenario{}.}
        \label{fig:balance_map_heat}
    \end{subfigure}
\end{figure}

\begin{figure}[ht!]
    \centering
    \begin{subfigure}{0.5\textwidth}
        \includegraphics[width=\linewidth]{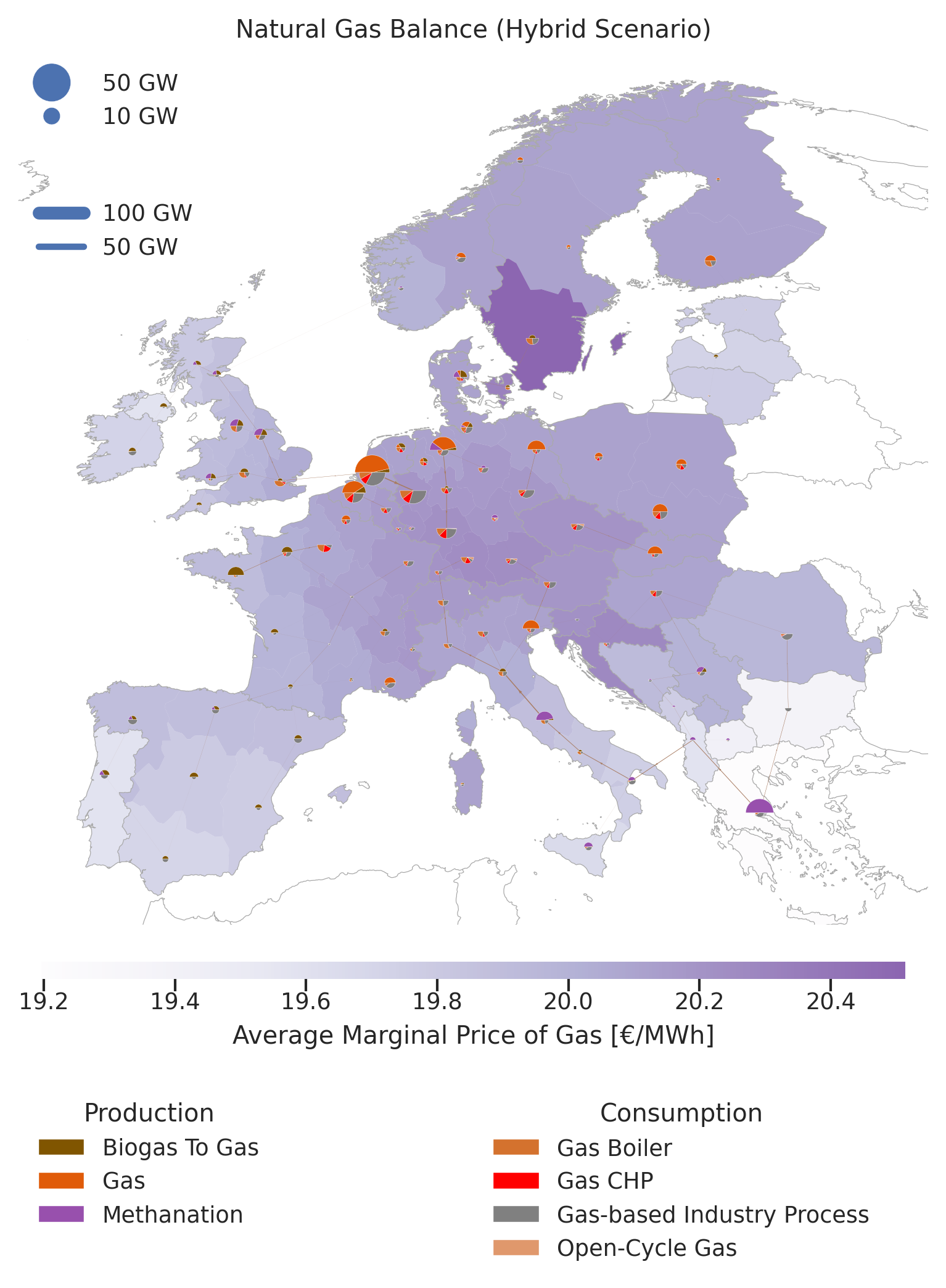}
        \caption{Optimal supply, consumption, transport and prices of gas in the \hybridscenario{}.}
        \label{fig:balance_map_gas}
    \end{subfigure}%
    \begin{subfigure}{0.5\textwidth}
        \includegraphics[width=\linewidth]{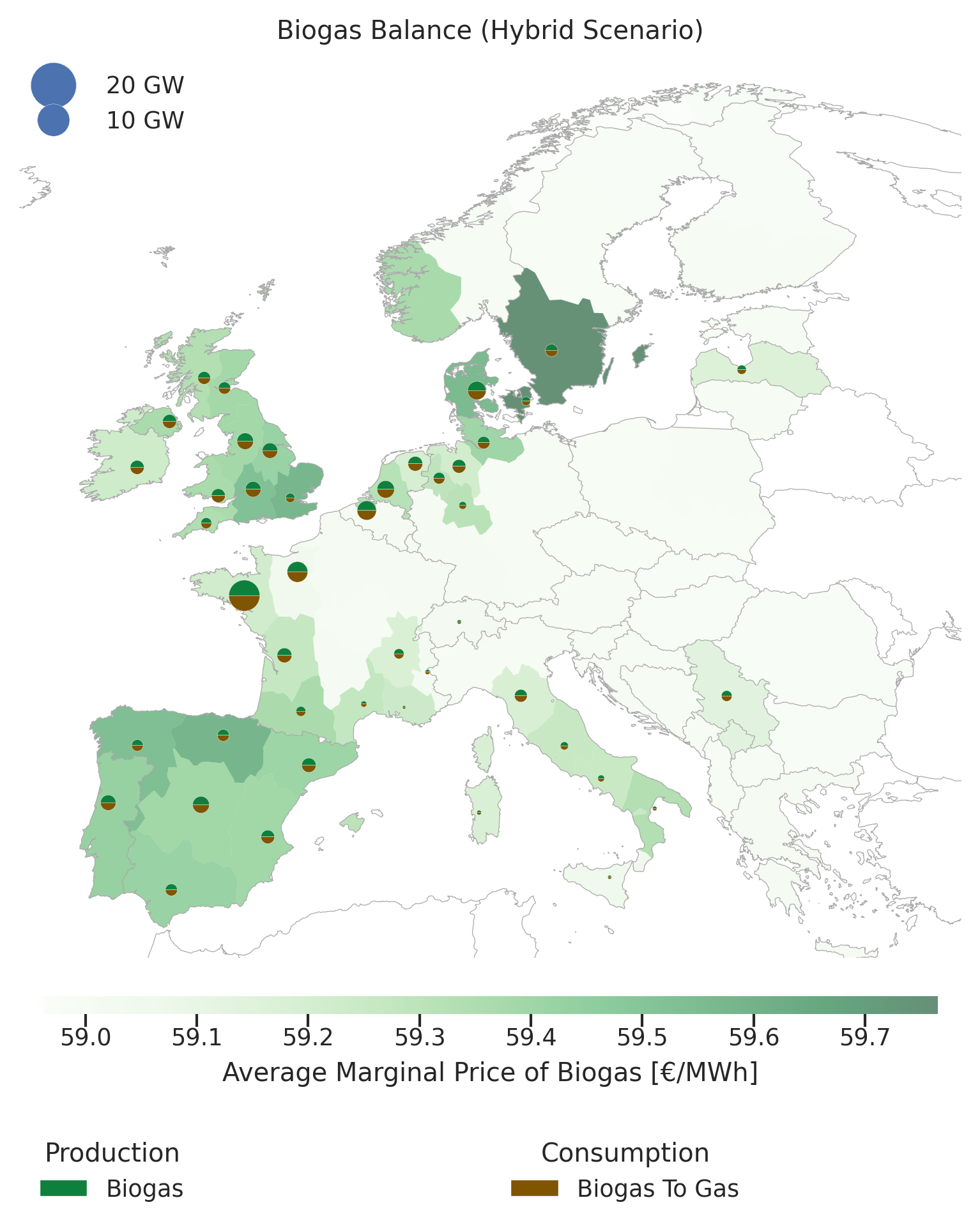}
        \caption{Optimal supply, consumption, transport and prices of biogas in the \hybridscenario{}.}
        \label{fig:balance_map_biogas}
    \end{subfigure}
\end{figure}

\clearpage
\subsection{Cost comparison across net-zero scenarios}
\label{sec:cost_comparison}

The following figures show further comparisons of total system costs and cost contributions of single technologies across the models and scenarios.

\begin{figure}[h!]
    \centering
    \includegraphics*[width=0.4\linewidth]{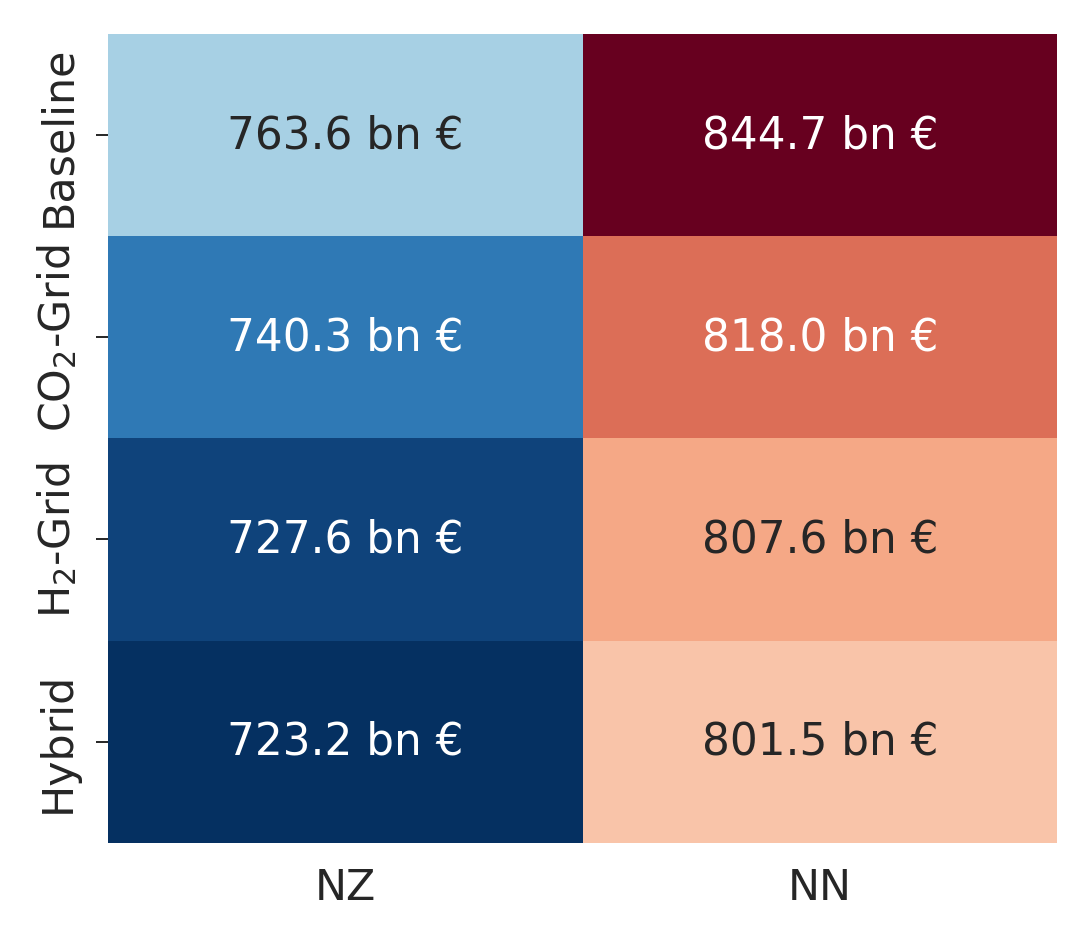}
    \caption{Total annual system cost for the different network models and \carbon{} targets, net-zero (NZ) left, net-negative (NN) right.}
    \label{fig:objective_heatmap}
\end{figure}

\begin{figure}[h!]
    \centering
    \begin{subfigure}{.5\textwidth}
        \centering
        \includegraphics[width=\linewidth]{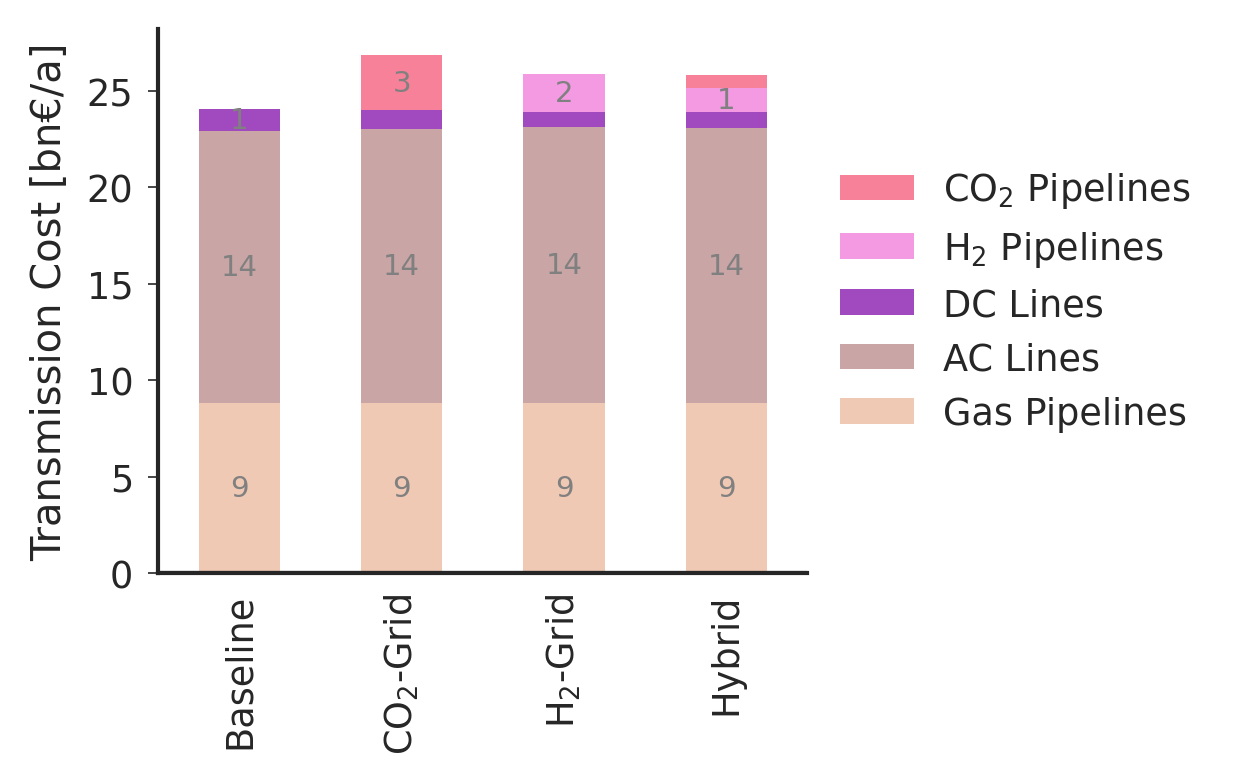}
        \caption{}
        \label{fig:cost_bar_transmission}
    \end{subfigure}%
    \begin{subfigure}{.5\textwidth}
        \centering
        \includegraphics[width=\linewidth]{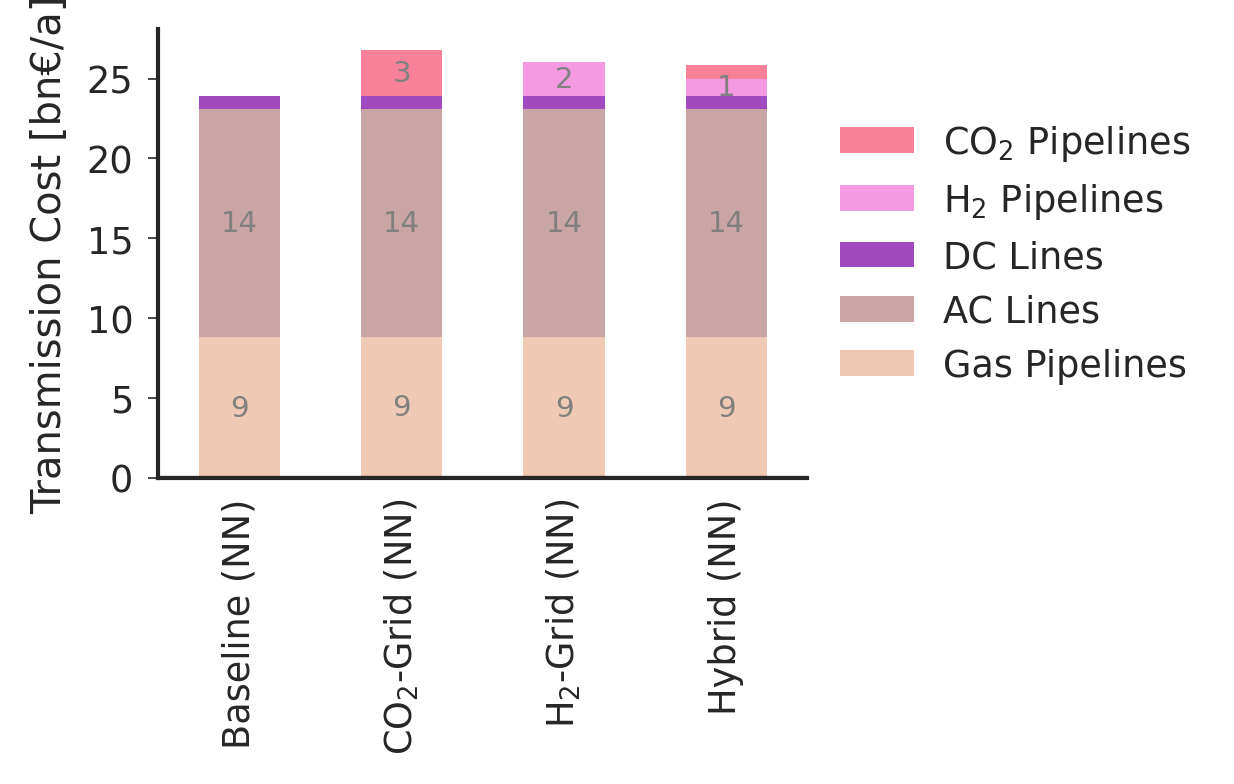}
        \caption{}
        \label{fig:cost_bar_transmission_nn}
    \end{subfigure}
    \caption{Annual transmission system costs for different models and scenarios for net \carbon{} neutrality (a) and net \carbon{} negative (NN) scenario (b).}
\end{figure}

\begin{figure}[ht!]
    \centering
    \includegraphics[width=\linewidth]{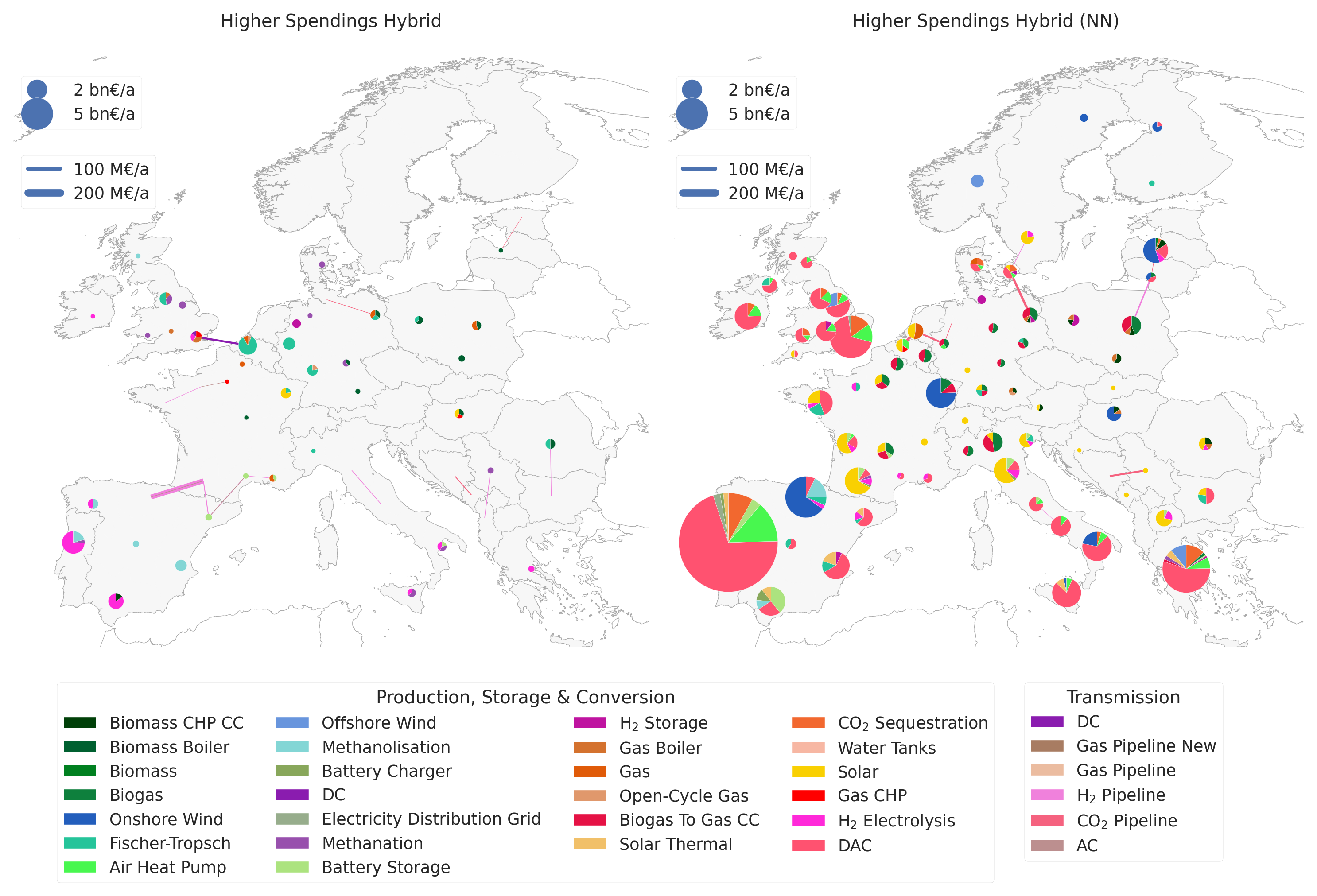}
    \caption{Difference in regional costs between when tightening the emission target in the \hybridscenario{} from \carbon{} net-neutrality to net-removal. The left figure shows higher spending per technology and region and transport system net-neutrality case, the right shows higher spending net-negative (NN) case.}
    \label{fig:cost_map_difference_full_nn}
\end{figure}

\begin{figure}[ht!]
    \centering
    \includegraphics[width=\linewidth]{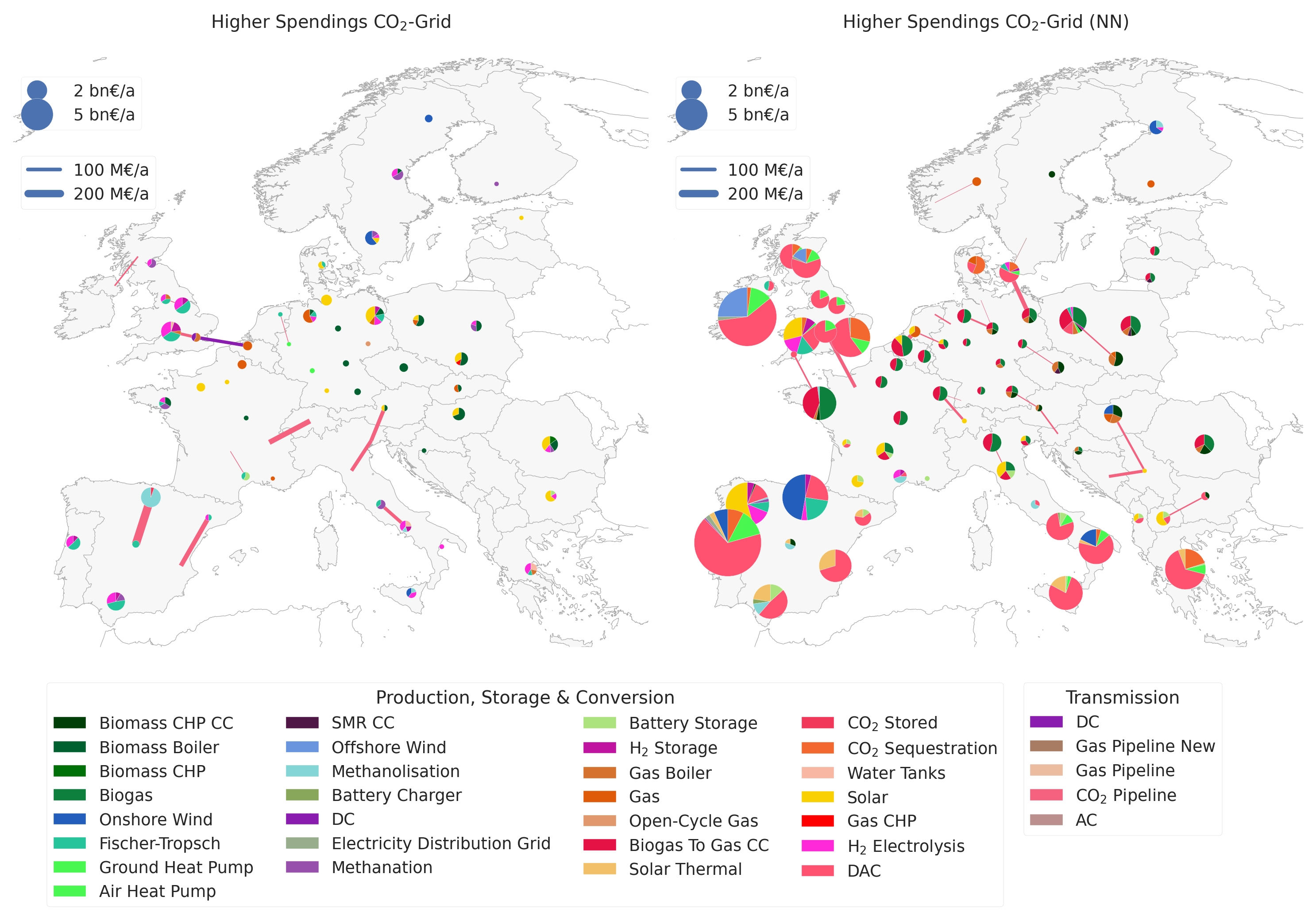}
    \caption{Difference in regional costs between when tightening the emission target in the \carbonscenario{} from \carbon{} net-neutrality to net-removal. The left figure shows higher spending per technology and region and transport system net-neutrality case, the right shows higher spending net-negative (NN) case.}
    \label{fig:cost_map_difference_co2_nn}
\end{figure}

\begin{figure}[ht!]
    \centering
    \includegraphics[width=\linewidth]{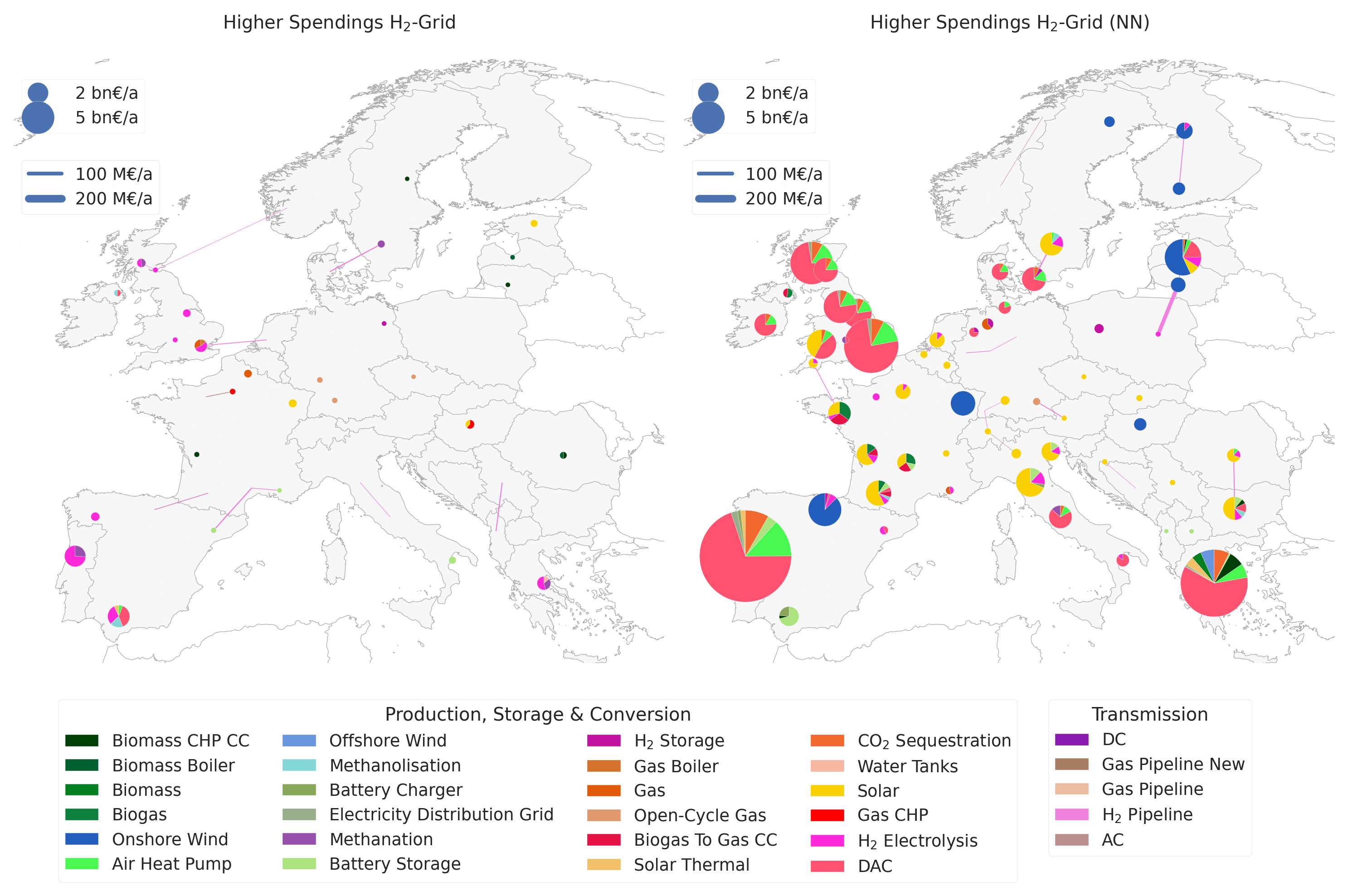}
    \caption{Difference in regional costs between when tightening the emission target in the \hydrogenscenario{} from \carbon{} net-neutrality to net-removal. The left figure shows higher spending per technology and region and transport system net-neutrality case, the right shows higher spending net-negative (NN) case.}
    \label{fig:cost_map_difference_h2_nn}
\end{figure}

% \begin{figure}[ht!]
%     \centering
%     \includegraphics[width=\linewidth]{difference/emission-reduction-0.1-co2-only/figures/90_nodes/cost_map.png}
%     \caption{Difference in regional costs between the \hydrogengrid{} and \hybridscenario{} assuming net \carbon{} neutrality. The left figure shows higher spending per technology and region and transport system in the \hydrogenscenario{}, and the right shows higher spending in the \hybridscenario{}.}
%     \label{fig:cost_map_difference}
% \end{figure}

% \begin{figure}[ht!]
%     \centering
%     \includegraphics[width=\linewidth]{difference/net-negative-0.1-h2-only-full/figures/90_nodes/cost_map.png}
%     \caption{Difference in regional costs between the \hydrogengrid{} and \hybridscenario{} in the net \carbon{} negative (NN) scenario. The left figure shows higher spending per technology and region and transport system for the \hydrogenscenario{}, and the right shows higher spending in the \hybridscenario{}.}
%     \label{fig:cost_map_difference_nn}
% \end{figure}

\clearpage
\subsection{Operation in Net \carbon{} Removal Scenarios}
\label{sec:operation_nn}
The following figures display the optimal operation of the hydrogen and carbon sector for all models in the net \carbon{} removal scenario.

\begin{figure}[ht!]
    \centering
    \begin{subfigure}{.5\textwidth}
        \centering
        \includegraphics[width=\linewidth]{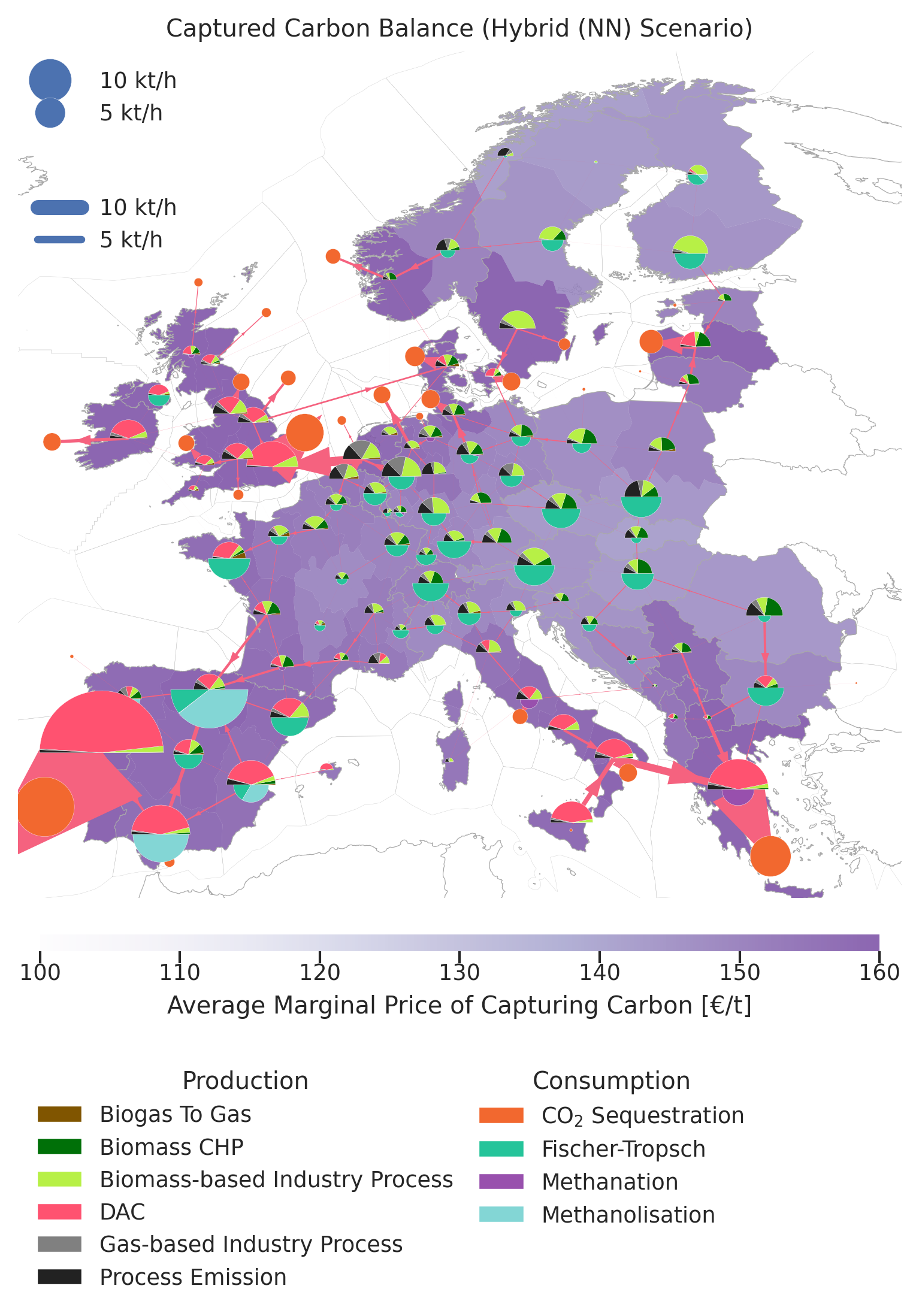}
        \label{fig:balance_map_carbon_full_nn}
    \end{subfigure}%
    \begin{subfigure}{.5\textwidth}
        \centering
        \includegraphics[width=\linewidth]{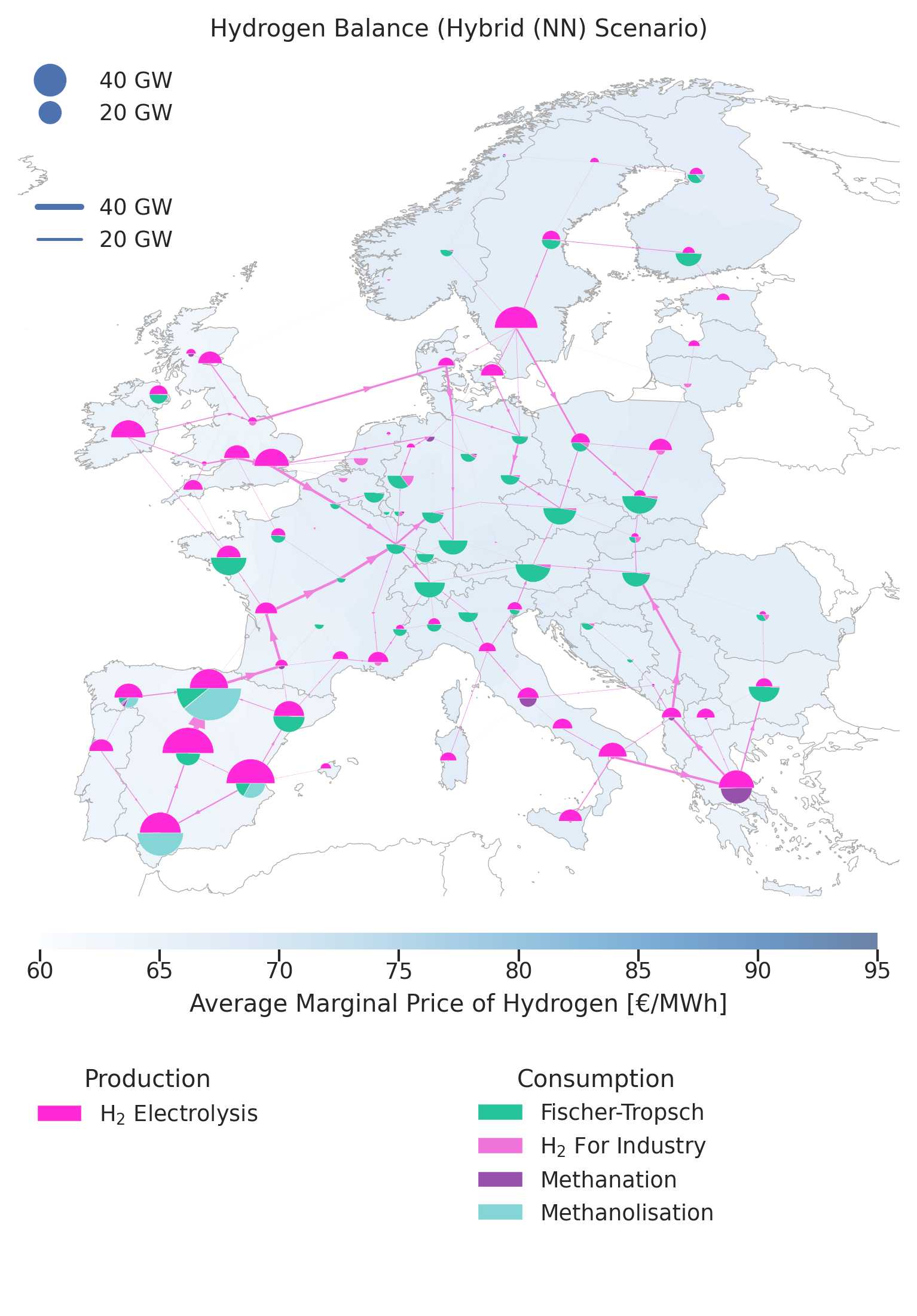}
        \label{fig:balance_map_hydrogen_full_nn}
    \end{subfigure}
    \caption{Optimal operation, flows and prices of the carbon (left) and hydrogen (right) sectors for the \hydrogenscenario{} in the net \carbon{} removal scenario. For each region, upper semicircles show the average production per technology, lower semicircles the consumption, and colors the average marginal prices. Carbon Sequestration offshore is drawn in full circles. Lines and arrows show the interregional transportation. With the tightened emission target, the model decreases the \hydrogengrid{} layout in comparison to the net carbon neutrality scenario and increases \carbon{} transport from inland point sources to the sequestration sites.
    }
    \label{fig:balance_maps_full_nn}
\end{figure}

\begin{figure}[ht!]
    \centering
    \begin{subfigure}{.5\textwidth}
        \centering
        \includegraphics[width=\linewidth]{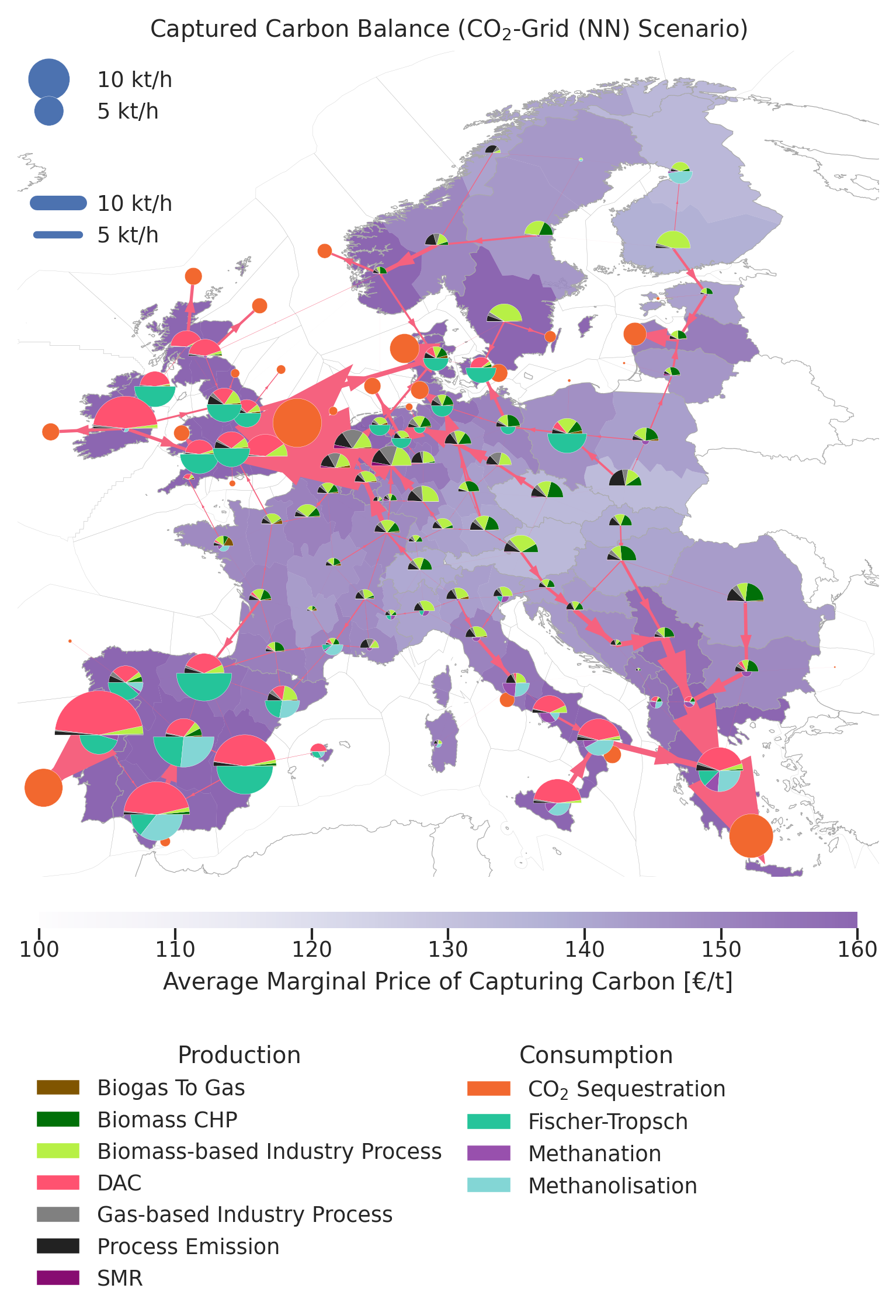}
        \label{fig:balance_map_carbon_co2_nn}
    \end{subfigure}%
    \begin{subfigure}{.5\textwidth}
        \centering
        \includegraphics[width=\linewidth]{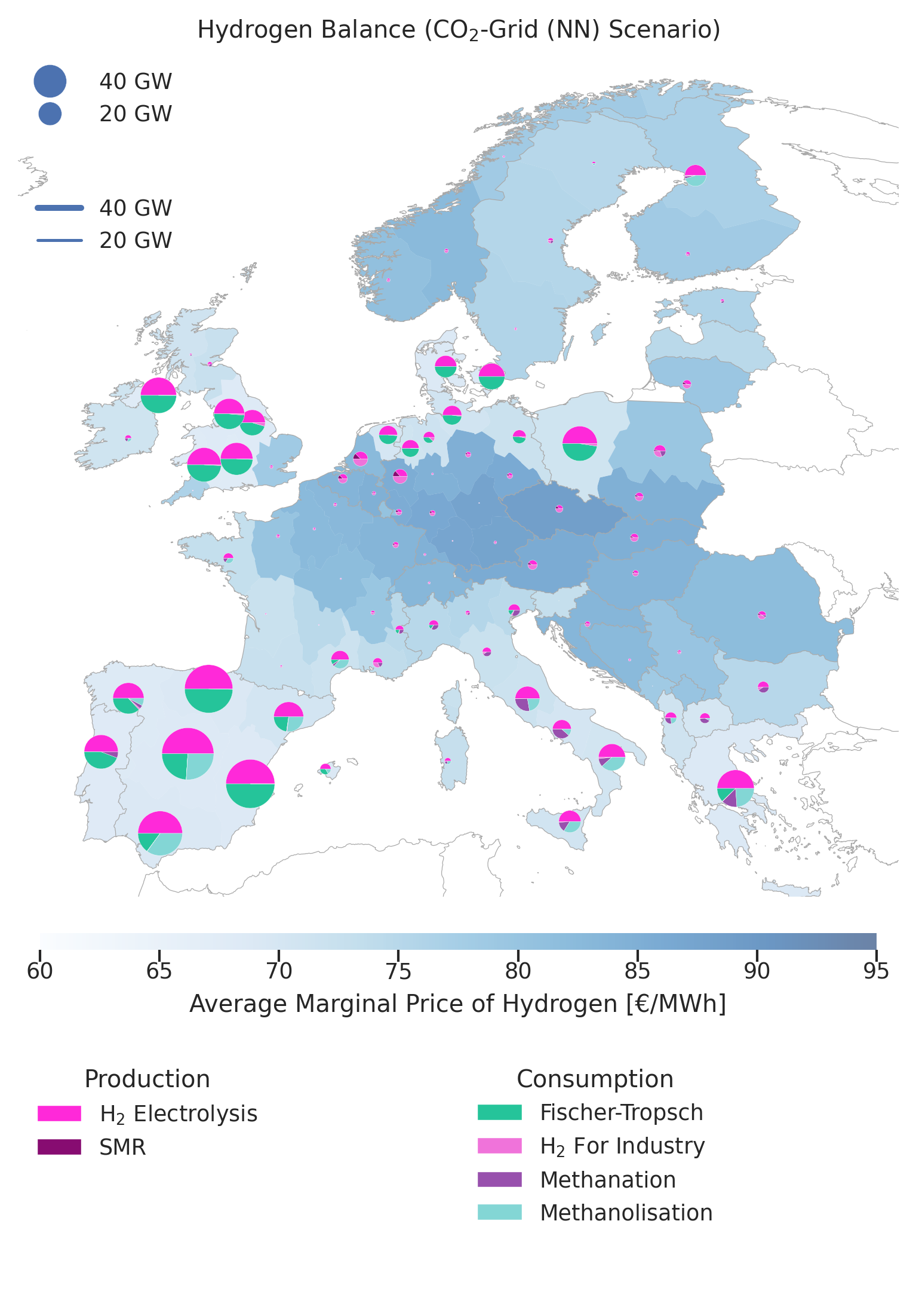}
        \label{fig:balance_map_hydrogen_co2_nn}
    \end{subfigure}
    \caption{Optimal operation, flows and prices of the carbon (left) and hydrogen (right) sectors in the \carbonscenario{} for net-negative emission targets. For each region, upper semicircles show the average production per technology, lower semicircles the consumption, and colors the average marginal prices. Carbon Sequestration offshore is drawn in full circles. Lines and arrows show the interregional transportation. The \carbongrid{} reveals a fundamentally different layout than with the net carbon neutrality target.
    }
    \label{fig:balance_maps_co2_nn}
\end{figure}

\begin{figure}[ht!]
    \centering
    \begin{subfigure}{.5\textwidth}
        \centering
        \includegraphics[width=\linewidth]{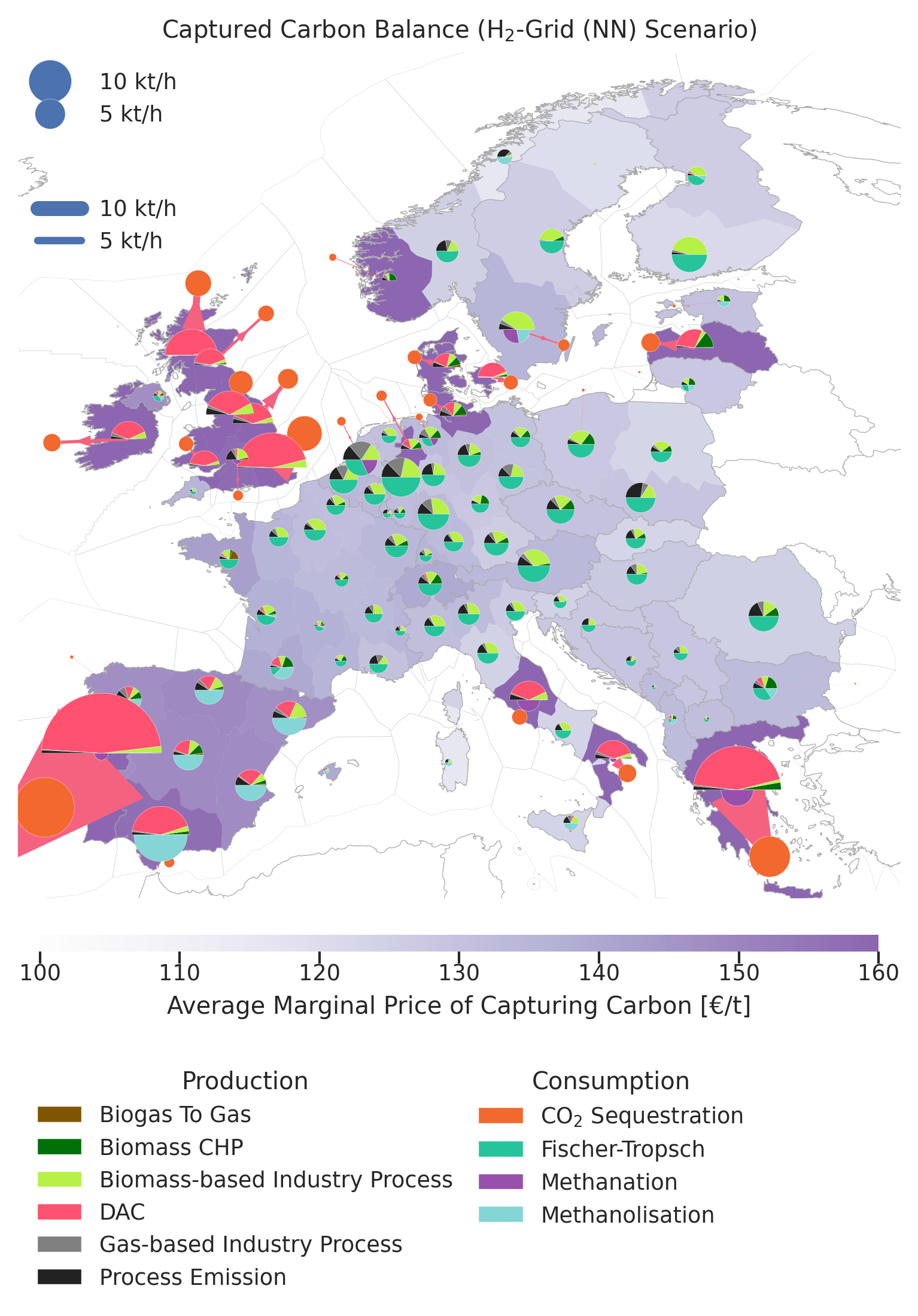}
        \label{fig:balance_map_carbon_h2_nn}
    \end{subfigure}%
    \begin{subfigure}{.5\textwidth}
        \centering
        \includegraphics[width=\linewidth]{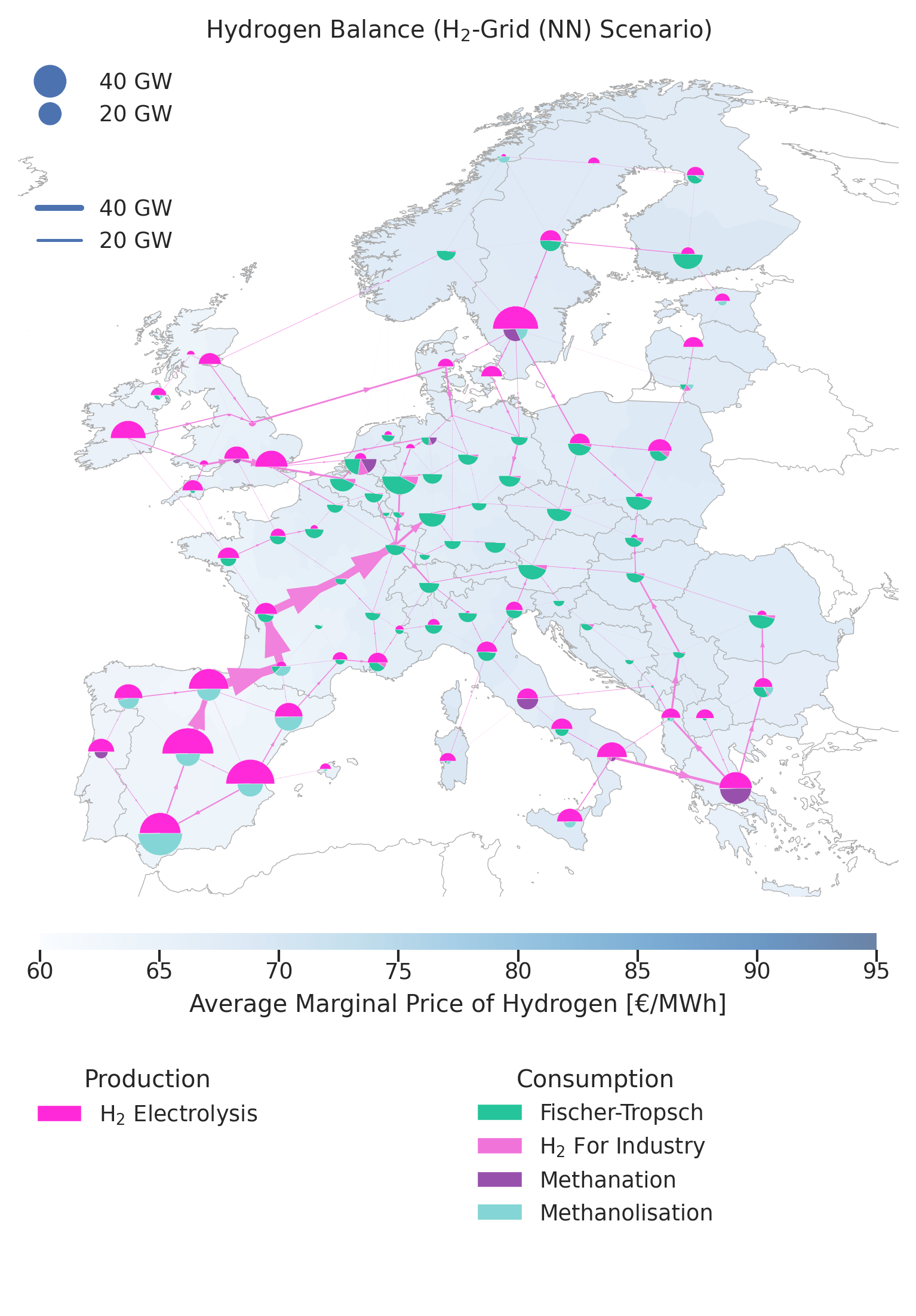}
        \label{fig:balance_map_hydrogen_h2_nn}
    \end{subfigure}
    \caption{Optimal operation, flows and prices of the carbon (left) and hydrogen (right) sectors in the \hydrogenscenario{} for net-negative emission targets. For each region, upper semicircles show the average production per technology, lower semicircles the consumption, and colors the average marginal prices. Carbon Sequestration offshore is drawn in full circles. Lines and arrows show the interregional transportation. With the tightened emission target, the \hydrogengrid{} expands the \hydrogengrid{} layout from the net carbon neutrality target results while increasing DAC facilities at the coast.
    }
    \label{fig:balance_maps_h2_nn}
\end{figure}

% \clearpage
% \subsection{Inter-sectoral \carbon{} flows}

% The following figure shows Sankey Diagrams of the carbon flows between the sectors.

% \begin{figure}[h!]
%     \centering
%     \includegraphics[width=.8\linewidth]{full/figures/90_nodes/sankey_diagramm.png}
%     \caption{Sankey diagram of the carbon flows in the \hybridscenario{} assuming net \carbon{} neutrality.}
%     \label{fig:sankey_diagramm}
% \end{figure}

% \begin{figure}
%     \centering
%     \includegraphics[width=.8\linewidth]{net-negative-0.1/full/figures/90_nodes/sankey_diagramm.png}
%     \caption{Sankey diagram of the carbon flows in the \hybridscenario{} with a net \carbon{} removal target.}
%     \label{fig:sankey_diagramm_nn}
% \end{figure}

\end{document}